\documentclass[12pt]{article}
\usepackage{graphicx}
\usepackage{epsfig}
\begin{document}
\topskip 2cm
\begin{titlepage}
\rightline{ \large{ \bf TTP00-16} }
\rightline{ \large{ \bf hep-ph/0008262} }
\begin{center}
{\large\bf Four Pion Final States with Tagged Photons \\
 at Electron Positron
Colliders} \\
\end {center}
\vspace{2.cm}
\begin{center}
{\large {\bf H.~Czy{\.z}\ $^{a}$ , J.H.~K{\"u}hn$^{b}$ }
   } 

\begin{itemize}
\item[$^a$] 
             {\sl Institute of Physics, University of Silesia,
             ul. Uniwersytecka 4, 
             PL-40007 Katowice, Poland }
\item[$^b$] 
             {\sl Institut f{\"u}r Theoretische Teilchenphysik, 
              Universit{\"a}t Karlsruhe, D-76128 Karlsruhe, Germany }

\end{itemize}

\end{center}
\noindent
e-mail: {\tt  czyz@us.edu.pl\\ 
\hspace*{1.3cm} johann.kuehn@physik.uni-karlsruhe.de \\ } 
\vspace{.5cm}

\begin{center}
\begin{abstract}

A Monte Carlo generator has been constructed to
simulate the reaction $e^+e^- \to \gamma + 4 \pi$, where the photon is
assumed to be observed in the detector.
Only initial state radiation is considered.
Additional
collinear photon radiation has been incorporated with the technique of
structure functions. Predictions are presented for cms energies of 1GeV,
3GeV and 10GeV, corresponding to the energies of DA\(\Phi\)NE, BEBC and
of $B$-meson factories.
The event rates are sufficiently high
to allow for a precise measurement of $R(Q^2)$ in the region of $Q$
between approximately 1GeV and 2.5GeV.
For the construction of the program we employ isospin relations
between the amplitudes
governing $\tau$ decays into four pions and electron positron
annihilation into four pions.
Estimates of the kinematic breaking of these isospin
relations as a consequence of the $\pi^-$ -- $\pi^0$ mass difference
 are given.

\end{abstract}
\end{center}

\vfill
\end{titlepage}
\pagestyle{plain} \pagenumbering{arabic} 
\newcommand{\Eq}[1]{Eq.(\ref{#1})} 
\newcommand{\labbel}[1]{\label{#1}} 
\newcommand{\be}{\begin{equation}}
\newcommand{\ee}{\end{equation}}
\newcommand{\ba}{\begin{eqnarray}}
\newcommand{\ea}{\end{eqnarray}}

\section{Introduction.}

 The precise determination of the cross section for electron positron
annihilation into hadrons over a large energy range is one of the
important tasks of current particle physics. The results are relevant
for the analysis of electroweak precision measurements which are
affected by the running of the electromagnetic coupling from the
Thompson limit up to $M_Z$. Also the interpretation of the anomalous
magnetic moment of the muon depends critically on these data. Last not
least the measurement of the energy dependence of R(s) is one of the
gold plated tests of QCD and allows for a precise determination of the
strong coupling constant.

Depending on the energy region different techniques for the measurement
of $R(s)$ have been applied up to date. At low energies, say from the
two pion threshold up to roughly two GeV, exclusive channels are
collected separately, for higher energies inclusive measurements start
to become dominant. For energies below $m_\tau$ isospin invariance and
CVC have traditionally been used to predict $\tau$ decays from electron
positron annihilation \cite{Sakur,Tsai,Gil,KS}. Clearly this
strategy can be inverted \cite{K1} pending irreducible uncertainties
from isospin violation and radiative corrections \cite{K2}.
At high energies a multitude of final states is present and only
inclusive measurements have been performed. 

To cover a large range of
energies, results from many different experiments and colliders have to
be combined, and energy scans have to be performed to obtain the full
energy dependence. An attractive alternative is provided by the
upcoming $\Phi-$ and $B-$ meson factories which operate at large
luminosities, albeit at fixed energies. Events with radiated tagged
photons give access to a measurement of $R$ over the full range of
energies, from threshold up to the CMS energy of the collider. For
events with tagged photons the invariant mass of the recoiling hadronic
system is fixed by the photon energy which provides an
important kinematic constraint. The usage of photons  observed at
extremely  small angle with respect to the beam has been investigated in
\cite{Kuraev1,Kuraev2}. In this case final state radiation as background is
practically irrelevant. However, in practice photon detectors
do not cover this region. At large angles
  a careful study of
initial versus final state radiation has to be performed.

To arrive at reliable predictions including angular and energy cutoffs
as employed by realistic experiments, a Monte Carlo generator 
 is indispensable. For hadronic states with
invariant masses below two or even three GeV it is desirable to simulate
the individual exclusive channels with two, three up to six mesons i.e.
pions, kaons, etas
 etc. which requires a fairly detailed parametrization of the
various form factors.

In principle initial and final state radiation would be required for the
complete simulation. Such a program has been constructed for the two
pion case \cite{BKM}. There it is demonstrated that suitably chosen
configurations, namely those with hard photons at small angles relative
to the beam and well separated from the pions, are dominated by initial
state radiation. In fact, this separation is possible \cite{Graziano} even
when operating the $\phi$ factory DA\(\Phi\)NE on top of the $\phi$ resonance
where direct radiative $\phi$ decays cannot be ignored.  

In the present paper we continue this project with the construction of a
generator for the radiative production of the four pion final state,
including the $\omega (\to3\pi) \pi$ channel. This mode contributes a
large fraction of the rate with invariant masses
  between one and two GeV. The
energy region between 1.5 GeV and 2.5 GeV is difficult to access 
directly with
current electron positron colliders. At the same time the experimental
uncertainties are relatively large. This motivates the special effort
devoted to this range.

The Monte Carlo program discussed in the present paper is
 specifically devoted to cms energies up to approximately 10 GeV.
It is constructed in a modular form such that the
parametrization of the hadronic matrix element can easily be replaced by
a more elaborate version. Different final states with three, four
or five pions or kaons can be included. The present
parametrization of the hadronic matrix element follows closely the form
suggested in \cite{Fink}, correcting only some minor deficiencies.
 The four pion amplitude is assumed to be dominated by
 \(\rho\prime \to \pi a_1 \) plus a direct coupling 
 \(\rho\prime \to \rho\pi\pi\) and exhibits the proper behavior in the chiral
 limit.

The plan of this paper is as follows: In the next section the
formalism for the decomposition of the differential cross section into a
leptonic and a hadronic tensor is presented. Results for partially
integrated distributions are recalled which can easily be used to arrive
at simple estimates for the rates. In section 3 isospin relations are
derived between the amplitudes for four pion production from a virtual photon 
and those accessible in $\tau$ decay. The relations between these four
matrix elements contain the well known identities between the
corresponding rates and provide in addition important constraints on
the differential distributions. In section 4 the influence of the
 \(\pi^0-\pi^\pm\) mass difference on the relations obtained in section 3 
is discussed.
The ingredients of the ansatz for the
hadronic amplitude  together with a comparison between the model
prediction and the data for a variety of distributions are presented in
section 5, the complete description of the hadronic amplitude
 with all the model parameters is
collected in the appendix. A description of the Monte Carlo generator is
given in section 6, together with a few characteristic distributions. In
particular we investigate the influence of angular cuts and the addition of
collinear radiation. Section 7
contains a brief summary and the conclusions.
\section{The radiative return.}

\begin{figure}[ht]
\begin{center}
\epsfig{file=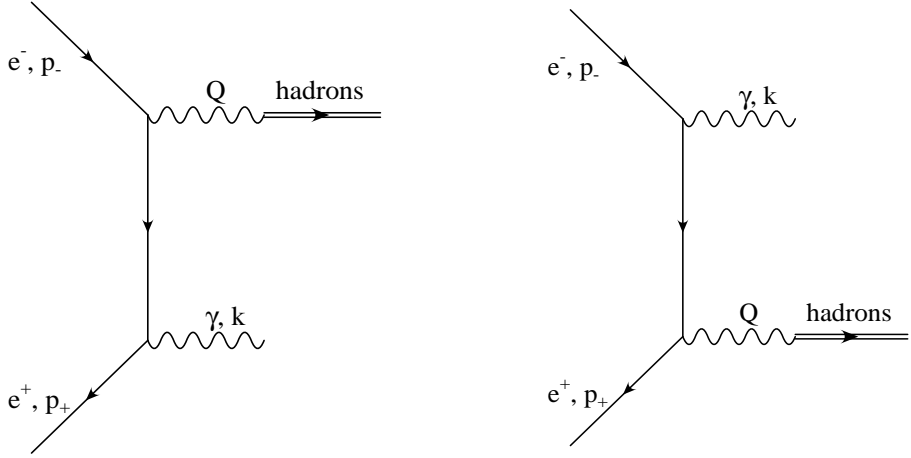,width=12cm,height=6cm}
\end{center}
\caption{Diagrams contributing to the process \(e^+e^- \to \gamma+ hadrons\)
  (only initial state radiation included).
}
\label{f1}
\end{figure}

  Hard photons observed at small angles relative to the electron or
positron beam and at the same time well separated from charged particles
in the final state can be used to reduce the effective center of mass
energy at electron positron colliders. Performing a detailed analysis of
the angular and energy distributions for the $\gamma \pi^+ \pi^-$ final
state it has been shown that initial and final state radiation can be
reasonably well separated \cite{BKM,Graziano}. For the four pion case we
therefore restrict the discussion to initial state radiation only. The
matrix element for the production of an arbitrary hadronic final state
corresponding to the diagrams Fig.\ref{f1} are given by

\begin{eqnarray}
 {\cal M} = i \ e^3 \ \bar v\left(p_{+}\right) 
 \left[ \gamma^{\nu} \frac{1}{p_{-}{\kern-12pt}/ - k {\kern-5pt}/ - m}
  \epsilon^{*}{\kern-10pt}/\left(k \right)
 + \epsilon^{*}{\kern-10pt}/\left(k \right) 
    \frac{1}{k {\kern-5pt}/ - p_{+}{\kern-12pt}/ - m}\gamma^{\nu}
 \right] \frac{1}{Q^2}\ J^{em}_{\nu} \ .
\end{eqnarray}

\noindent
The matrix element of the hadronic current

\begin{eqnarray}
J^{em}_{\nu} \ \equiv \ J^{em}_{\nu}\left(q_1,...,q_n\right) \ \equiv \ 
 <h(q_1),...,h(q_n)|J^{em}_{\nu}\left(0\right)|0>
\end{eqnarray}

\noindent
has to be parameterized by form factors to be discussed below. For the
two pion case the current 

\begin{eqnarray}
J_{\nu}^{em,2\pi} 
 = \left(q_{\nu}^{+}-q_{\nu}^{-}\right)F_{2\pi}\left(Q^2\right)
\end{eqnarray}

\noindent
is determined by only one function, the pion
form factor \(F_{2\pi}\).

For the three pion case the matrix element of the electromagnetic
current is restricted by current conservation and negative parity to the
form

\begin{eqnarray}
 J_{\nu}^{em,3\pi} =
 \epsilon_{\nu\alpha\beta\gamma}\ q_{+}^{\alpha}q_{-}^{\beta}q_{0}^{\gamma}
  \ F_{3\pi}\left(q_{+},q_{-},q_{0}\right)
\end{eqnarray}

\noindent
and $F_{3\pi}$ is dominated by the $\omega$ resonance. The matrix element
for the four pion case will be discussed in sections 3 and 4.

The differential rate can be cast into a product of a leptonic and a
hadronic tensor and a corresponding separation of the phase space

\begin{eqnarray}
d\sigma = \frac{1}{8s} \ L_{\mu\nu} H^{\mu\nu}\  
    d\Phi_2(p_{+}+p_{-};Q,k)\  d\bar\Phi_n(Q;q_1,\dots,q_n) \ 
 \frac{dQ^2} {2\pi} \ , 
\labbel{SIG}
\end{eqnarray}

\noindent
where \(d\bar\Phi_n(Q;q_1,\dots,q_n)\) denotes the \(n\) body phase space
 with all statistical factors included.

The leptonic tensor $L_{\mu\nu}$ is process independent, the modeling
of hadronic physics enters the tensor $H_{\mu\nu}= J^{em}_\mu (J^{em}_\nu)^*$
only. The leptonic tensor is symmetric, and hence it is only the real
symmetric part of $H_{\mu\nu}$ which enters. It has the following form

\begin{eqnarray}
L^{\mu\nu} = 
  \frac{-4}{(kp_{-})(kp_{+})} 
  \left[ \ g^{\mu\nu} \  \left(
  p_{-}p_{+}  
 + \frac{(kp_{+})^2+ (kp_{-})^2}{Q^2} \right)
 + \ \ p_{+}^{\mu}p_{+}^{\nu}+p_{-}^{\mu}p_{-}^{\nu} \ \right] 
 \frac{1}{Q^2}
 \ .
\end{eqnarray}

The collinear and soft photon singularities proportional to $1/(1\pm
\cos\theta_\gamma)$ and $1/E_\gamma$ are evident from these expressions,
as well as the $1/Q^2$ enhancement for small $Q^2$. Integrating
the hadronic tensor $H_{\mu\nu}$ over the hadronic phase space one gets

\begin{eqnarray}
 \int \ J^{em}_\mu (J^{em}_\nu)^*  \ \ d\bar\Phi_n(Q;q_1,\dots,q_n) = 
 \frac{1}{6\pi} \left(Q_{\mu}Q_{\nu}-g_{\mu\nu}Q^2\right) \ R(Q^2)
\labbel{rr}
\end{eqnarray}
 
\noindent
where \(R(Q^2)\) is \(\sigma(e^+e^-\rightarrow hadrons)/\sigma_{point}\).

\begin{table}
\begin{center}
$$
\begin{array}{||c||c||c||c|c|c||}
\hline
&
&
&
\multicolumn{3}{|c|}{{\rm Event~rates}}\\
\hline
{\rm Collider }& \sqrt{s}
&{\rm Integrated~luminosity}, {\rm fb}^{-1} & \theta_{\rm min} = 5^\circ 
& \theta_{\rm min} = 7^\circ   &  \theta_{\rm min} = 10^\circ 
\\ \hline \hline
{\rm DA}\Phi{\rm NE} & 1.02 & 1 & 13\cdot 10^6 & 12 \cdot 10^6 
&  10 \cdot 10^6 \\ \hline \hline
B-{\rm factory} & 10.6 & 100 & 4 \cdot 10^6 & 3.5 \cdot 10^6 
&  3 \cdot 10^6 \\ \hline \hline
{B-\rm factory} & 10.6 & 100 & 2.7\cdot 10^6 & 2.3\cdot 10^6
&  2.0\cdot 10^6 \\ 
\hline
\end{array}
$$
\vspace*{0.5cm}
\caption{Estimated number of  radiative 
events $e^+e^- \to \ hadrons \ + \ \gamma$ for different center of mass
energies from \Eq{1}. In the first two rows by \(hadrons\) we mean
\(\pi^+\pi^-\) and 
the minimal photon energy is $0.1$ GeV. The third row is 
 obtained for a continuum
 contribution in the region
  \(2 \  {\rm GeV} < \sqrt{Q^2} < 3.7 \ {\rm GeV}\)
 assuming a constant \(R=2.4\).  }
\end{center}
\label{t1}
\end{table}

 The additional integration over the photon angles
 (the azimuthal angle is integrated over the full range and
the polar angle within \(\theta_{min} < \theta < \pi-\theta_{min}\) )
 leads to
  the differential distribution 

\begin{eqnarray}
Q^2 \frac {{\rm d}\sigma}{{\rm d} Q^2} = \frac
{4 \alpha^3}{3 s} R(Q^2)
\left \{ \frac {(s^2+Q^4)}{s(s-Q^2)} 
\log \frac {1+\cos \theta_{\rm min}}{1-\cos \theta_{\rm min}}
-\frac {(s-Q^2)}{s}\cos \theta_{\rm min} \right \} \ ,
\labbel{1}
\end{eqnarray}

\noindent
 which can be used to calculate the event rate observed for realistic photon
energy and angular cuts (see Tab.1).

\section{Isospin relations.}

The emphasis of this paper is towards hadronic final states consisting of
 four pions and a photon. Before entering a discussion of a model 
 dependent parametrization of the form factors (see section 5) we recall
 the constraints from isospin invariance. They relate the amplitudes
 of the \(e^+ e^- \to 2\pi^+2\pi^-\) and \(e^+ e^- \to \pi^+\pi^- 2\pi^0\)
 processes and those for \(\tau\) decays into \(\pi^-3\pi^0\) and
 \(\pi^+2\pi^-\pi^0\). The amplitude of the \(\tau\) decay into an arbitrary
 number of hadrons plus a neutrino is given by

\begin{eqnarray}
 {\cal M}_{\tau} = \ \ \frac{G_F}{\sqrt{2}} \ \cos\theta_c \ \
 \bar v\left(p_{\nu}\right) \gamma^{\alpha}\left(1-\gamma_5\right)
 u\left(p_{\tau}\right) \ \ J_{\alpha}^{-} \ ,
\end{eqnarray}

with 

\begin{eqnarray}
 &&J^{-}_{\alpha} \ \equiv \ J^{-}_{\alpha}\left(q_1,...,q_n\right) \ \equiv \ 
 <h(q_1),...,h(q_n)|J^{-}_{\alpha}\left(0\right)|0> \nonumber \\
 {\kern-30pt} {\rm and} \ \ \ \ \
 &&J^{-}_{\alpha}(0) = \bar d \ \gamma_\alpha \ u
 \ \ \ .
\end{eqnarray}
We use the same letter $J$ for the operator and its matrix element and
restrict our considerations to the Cabbibo allowed vector part of the
hadronic current.

\noindent
It leads to the differential distribution

\begin{eqnarray}
\frac{d\Gamma}{dQ^2} = 2 \ \Gamma_e \frac{\cos^2\theta_c}{m_{\tau}^2}
 \left(1-\frac{Q^2}{m_{\tau}^2}\right)^2 
 \left(1+2\frac{Q^2}{m_{\tau}^2}\right) R^{\tau}\left(Q^2\right) \ ,
\labbel{diff}
\end{eqnarray}

with 

\begin{eqnarray}
 \int \ J_\mu^{-} J^{-*}_\nu  \ \ d\bar\Phi_n(Q;q_1,\dots,q_n) = 
 \frac{1}{3\pi} \left(Q_{\mu}Q_{\nu}-g_{\mu\nu}Q^2\right) \ R^{\tau}(Q^2) \ .
\labbel{rrt}
\end{eqnarray}
Note the relative factor of 2 between the definitions in \Eq{rr} and 
 \Eq{rrt}.

Ignoring the issues of isospin breaking and radiative corrections,
the electromagnetic current can be decomposed into an isospin singlet piece
and a part transforming like the third component of an isospin triplet:

\ba
 J^{em} = \frac{1}{\sqrt{2}}\ \  J^3 + \frac{1}{3\sqrt{2}}\ \  J^{{\rm I}=0}
\ea

\noindent
 whereas the charged current generating \(\tau\) decays is given by

\ba
J^- = \frac{1}{\sqrt{2}} \ \ (J^1-i \ J^2) \ \ .
\ea

Final states with an even number of pions are produced through the isospin
 one part only, whence

\begin{eqnarray}
 \sqrt{2} \ \ J^{em}_{\mu}(2\pi) = J^{-}_{\mu}(2\pi)
 \labbel{rel0}
\end{eqnarray}

\noindent
and \(R(Q^2)=R^{\tau}(Q^2)\) for two pion final states.
 
 A similar relation for the four pion final state is easily obtained 
 as follows: the transition from the vacuum to four pions is mediated
 through a current \(\vec J_{\mu}\left(0\right)\) of the form

\ba
\vec J_{\mu}\left(0\right) =
 \frac{1}{4} \ \int d^4q_1 d^4q_2 d^4q_3 d^4q_4 \ \
   J_{\mu} \ 
   \left(\vec \pi_1\cdot\vec \pi_2 \right)\ \
 \left(\vec \pi_3
 \times \vec \pi_4 \right) \ ,
 \labbel{curr}
\ea

\noindent
where \(\vec\pi_i \equiv \vec\pi(q_i)\) denotes the pion field in the
 Cartesian basis. 
The letter $J$ is again used both for the operator and the function in the
integrand which corresponds essentially to the transition amplitude.
The function 
 \(J_{\mu} \equiv J_\mu(q_1,q_2,q_3,q_4) \) is symmetric (anti-symmetric)
 with respect to the interchange of \(q_1\) and \(q_2\) (\(q_3\) and \(q_4\)).
 The combination of pion fields relevant for the transition to four pions 
 with total charge -1 and 0 respectively is given by
\ba
 J^{-}_{\mu}\left(0\right) &=&
 \frac{1}{4} \ \int d^4q_1 d^4q_2 d^4q_3 d^4q_4 \ \
    J_{\mu}\cdot
 \left( \pi^+_1 \pi^-_2+ \pi^-_1 \pi^+_2
       + \pi^0_1 \pi^0_2\right) \left( \pi^-_3 \pi^0_4- \pi^-_4 \pi^0_3\right)
\nonumber \\
 J^{3}_{\mu}\left(0\right) &=&
 \frac{1}{4} \ \int d^4q_1 d^4q_2 d^4q_3 d^4q_4 \ \
    J_{\mu}\cdot
  \left( \pi^+_1 \pi^-_2+ \pi^-_1 \pi^+_2
       + \pi^0_1 \pi^0_2\right)\left( \pi^+_3 \pi^-_4- \pi^-_3 \pi^+_4\right)
 \ . 
\ea

Taking the matrix element of these operators between vacuum and the states

 \noindent
\(\langle \pi^+(p^+)\pi^-(p^-)\pi^0(p_1)\pi^0(p_2)|\) etc. one immediately
 arrives at

\ba
\langle \pi^+ \pi^- \pi_1^0 \pi_2^0 | J^3_{\mu} | 0 \rangle &=& 
J_{\mu}(p_1,p_2,p^+,p^-) \nonumber \\
\langle \pi^+_1 \pi^+_2 \pi^-_1 \pi^-_2 | J^3_{\mu} | 0 \rangle &=&
J_{\mu}(p_2^+,p_2^-,p_1^+,p_1^-) +
J_{\mu}(p_1^+,p_2^-,p_2^+,p_1^-) \nonumber \\
&&{\kern-15pt}+J_{\mu}(p_2^+,p_1^-,p_1^+,p_2^-)+
 J_{\mu}(p_1^+,p_1^-,p_2^+,p_2^-)\nonumber \\
\langle \pi^- \pi^0_1 \pi^0_2 \pi^0_3 | J^{-}_{\mu} | 0 \rangle &=& 
J_{\mu}(p_2,p_3,p^-,p_1)+
J_{\mu}(p_1,p_3,p^-,p_2)+J_{\mu}(p_1,p_2,p^-,p_3)
\nonumber \\
\langle \pi^-_1 \pi^-_2 \pi^+ \pi^0 | J^{-}_{\mu} | 0 \rangle &=& 
J_{\mu}(p^+,p_2,p_1,p^0)+
J_{\mu}(p^+,p_1,p_2,p^0)
 \labbel{rel} \ ,
\ea

\noindent
which connects \(\tau\) decay and electron positron annihilation.
It is clear from \Eq{rel}
 that only one matrix element, namely the one for \((+\ -\ 0\ 0)\) ,
 needs to be programmed and the remaining ones can be obtained by relabeling
 arguments. Interference terms between the two partitions \cite{Pais,PS,Gil}
  (3,1) and (2,1,1) are present in the differential distributions.
 For the integrated rates induced by the currents one obtains

\ba
 R\left(+\ - \ 0 \ 0\right) &=& \frac{1}{2} \ A 
 {\kern+18pt} \ \ \ ; \ \ \ 
 R^{\tau}\left(-\ - \ + \ 0\right) =  A + \frac{1}{2} \ B \nonumber \\
 R\left(+\ + \ - \ -\right) &=&  A \ +  \ B \ \ \ ; \ \ \ 
   R^{\tau}\left(-\ 0 \ 0 \ 0\right) {\kern+9pt}
 = \frac{1}{2}\left( A \ +  \ B\right) \ ,
\ea 

\noindent
with

\ba
 A &=& -\frac{2\pi}{Q^2}\int \ J^\mu\left(q_1,q_2,q_3,q_4\right)
             J^{*}_\mu \left(q_1,q_2,q_3,q_4\right)
    \ \ d\bar\Phi_4(Q;q_1,\dots,q_4)\nonumber \\
 B &=& -\frac{4\pi}{Q^2}\int \ {\rm Re}\left(J^\mu\left(q_1,q_2,q_3,q_4\right)
             J^{*}_\mu \left(q_1,q_3,q_2,q_4\right) \right)
    \ \ d\bar\Phi_4(Q;q_1,\dots,q_4) \ ,
\ea

\noindent
consistent with the familiar relations between \(\tau\) decays and
 \(e^+e^-\) annihilation into four pions:

\ba
 R^{\tau}\left(-\ 0 \ 0 \ 0\right) &=&  \frac{1}{2}\  R\left(+\ + \ - \ -\right)
\nonumber \\
 R^{\tau}\left(-\ - \ + \ 0\right) 
 &=& \frac{1}{2} \ R\left(+\ + \ - \ -\right) \ 
 + \ R\left(+\ - \ 0 \ 0\right) \ .
\labbel{CVC}
\ea
\section{The \(\pi^{\pm}\)-\(\pi^0\) mass difference 
 and the isospin relations.}

 All the relations obtained in the previous section are strictly
 applicable only
 in case all pions in the final states have the same mass, which is
 obviously not true. The relatively large (about 3.6\%) 
\(\pi^{\pm}\)-\(\pi^0\) mass difference will affect the 
 \(R(Q^2)\leftrightarrow R^\tau(Q^2)\) relation even if the relations
 \Eq{rel0} and \Eq{rel} still hold. The CVC hypothesis 
 and the assumption that 
transitions to
an even number of pions in the final state are described
by the iso-vector current
are well established experimentally \cite{EI}.
 It is thus natural to assume that the relations between currents hold
 and the question is up to what precision we can ignore the 
 \(\pi^{\pm}\)-\(\pi^0\) mass difference. This should give at least 
 an indication of the size of these "kinematic" isospin violations.
 We will address this issue
 here first for two pion states in some detail and subsequently
 indicate the size of possible effects for the four pion case.

 In order to estimate the size of kinematical isospin violations we 
 first start
 from the assumption that the cross section \(\sigma(e^+e^-\to\pi^+\pi^-)\)
 is well measured and the squared form factor is extracted through
 
 \ba
   |F(Q^2)|^2 = \frac{3 Q^2}{\pi \alpha^2 \beta_{\pi}^3} 
 \ \sigma(e^+e^-\to\pi^+\pi^-)(Q^2) \ ,
 \ea

\noindent
  \( \left( \beta_{\pi} 
 = \left(1 - 4 m_{\pi^-}^2 / Q^2 \right)^{1/2}  \right)\)
 and used to predict the \(\tau^-\to\nu_{\tau}\pi^-\pi^0\) decay rate
 (We ignore the electroweak correction factor \(S_{ew}\simeq 1.019.)\).
The size of the corrections depends critically on the details of 
the assumptions.

If the form factor and the form of the current

\ba
 J^{em}_\mu(2\pi) = \sqrt{2} \ J^-_\mu(2\pi) = (q_{1,\mu}-q_{2,\mu})
 \ F(Q^2) 
 \labbel{cur1}
\ea 

\noindent
 remain unchanged, the integral for the \(\tau\) rate is given by

\vskip 1 cm

\ba
 &&{\kern-40pt}\frac{\Gamma(\tau^-\to \nu_\tau\pi^-\pi^0)}{\Gamma_e} \ \
  = \nonumber \\ 
   &&\frac{\cos^2\theta_c} { 8 m_{\tau}^2}
 \int dQ^2
  \ \left(1-\frac{Q^2}{m_{\tau}^2}\right)^2 
 \left[ \beta^3_{-0}\left(1+2\frac{Q^2}{m_{\tau}^2}\right) 
  +  3   \ \beta_{-0}\frac{(m_{-}^2-m_0^2)^2}{(Q^2)^2}
 \right] \ \ |F(Q^2)|^2 \ , 
 \labbel{CVCf}
\ea 

with
 \ba 
  \beta_{-0} \equiv \beta(Q^2,m_{-}^2,m_{0}^2)= 
 \left[\left(1-\frac{(m_{-}+m_0)^2}{Q^2}\right)
 \left(1-\frac{(m_{-}-m_0)^2}{Q^2}\right)\right]^{1/2} \ .
 \ea

\noindent
The second term is due to an S-wave contribution 
 and can be eliminated
 by replacing \Eq{cur1} by

\ba
 \sqrt{2} \ J^-_\mu(2\pi) = \left(q_{1,\mu}-q_{2,\mu} 
   - \frac{q_{1,\mu}+q_{2,\mu}}{Q^2}\ Q.(q_1-q_2)\right) \ F(Q^2) \ .
\ea 

\noindent
Numerically the contribution to the integral of this latter term is tiny
 - nevertheless we shall adopt the second choice.

This purely kinematic modification\footnote{This is the strategy 
 adopted in the generator TAUOLA \cite{tauola} which was written 
when data were not precise enough to require the incorporation of 
 mass corrections
 in the form factor.} raises the prediction for the \(\tau\) decay rate
 by 0.86\%. It seems, however, plausible, that the energy dependent width
 of th \(\rho\)- meson, which is present in the form factor, has to be modified
 accordingly\footnote{J.K. thanks M. Davier for drawing his attention
 to this point.} 
 leading to an effective increase of \(\Gamma_{\rho}\) by 0.74\%.
The two effects nearly compensate in the
 integral. Hence the relation between \(\tau^- \to \nu_{\tau} \pi^- \pi^0 \) 
 partial decay width and the \(e^+e^-\to\pi^+\pi^-\) cross section
 based on

\ba
 &&{\kern-40pt}\frac{\Gamma(\tau^-\to \nu_\tau\pi^-\pi^0)}{\Gamma_e} \ \
  = \nonumber \\
   &&\frac{3\cos^2\theta_c} {2\pi \alpha^2 m_{\tau}^2}
 \int dQ^2
 \ \ Q^2 \ \left(1-\frac{Q^2}{m_{\tau}^2}\right)^2 
 \left(1+2\frac{Q^2}{m_{\tau}^2}\right) 
 \ \sigma(e^+e^-\to\pi^+\pi^-) 
 \labbel{CVCi}
\ea 

\noindent
 would only be corrected by 0.06\%.

\begin{figure}[htbp]
\epsfig{figure=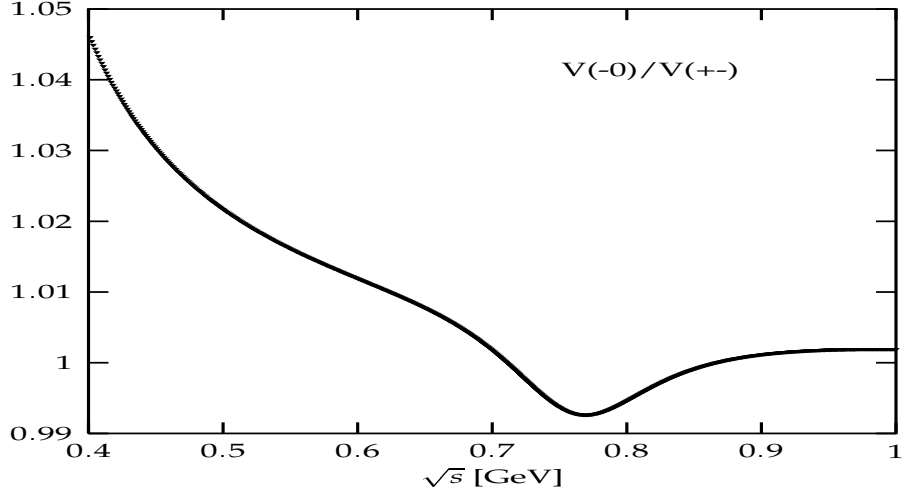, width=0.75\textwidth,height=6.5cm}
\caption{The ratio 
 of the two spectral functions (see text for explanation).
}
\label{fm}
\end{figure}

 However, as shown in Fig.\ref{fm}, a sizable \(Q^2\) dependence
 of the ratio of the two spectral functions
 (\(V(Q^2) \sim |F(Q^2)|^2 \beta^3\) )
 is expected, with a reduction approximately 0.74 \% close to the peak
 of the \(\rho\) resonance and enhancements at the tails. 
\begin{table}
\begin{center}
$$
\begin{array}{|c|c|c|}
\hline
  
& {\rm only} \  \beta^3  \ {\rm modified }
&   \beta^3 \  {\rm and}  \ \Gamma_{\rho} \ {\rm modified }\ \\
\hline
 0.28\  < \ E \ [GeV] \ < 0.81
&
&  \\
 a_{\mu}
&1.0163
&1.0088 \\
 \Delta \alpha(M_Z)
&1.0116   
&1.0027 \\
\hline
 0.32\  < \ E \ [GeV] \ < 1.777
&
&  \\
 a_{\mu}
&1.0130
&1.0058 \\
 \Delta \alpha(M_Z)
&1.0096
&1.0016 \\
\hline
\end{array}
$$
\vspace*{0.5cm}
\caption{
Kinematic correction factors for the predictions of \(a_{\mu}\)
 and \(\Delta\alpha\) from \(\tau\) data. 
 }
\end{center}
\label{tspec}
\end{table}

 At present, however, \(\tau\) data provide an important input for the 
 prediction of the QED coupling at the scale of \(M_Z\) and the hadronic
 contribution to g-2 \cite{K1,ADH}. Let us, for the moment, assume that
 the aforementioned kinematic effects are indeed present. If the 
 \(e^+e^-\) cross section is deduced from \(\tau\) data through \Eq{CVCc}

\begin{eqnarray}
&&{\kern-40pt}\frac{1}{\Gamma_e} \ \
\frac{d\Gamma(\tau^-\to \nu_\tau\pi^-\pi^0)}{dQ^2} = \nonumber \\
   &&\frac{3\cos^2\theta_c} {2\pi \alpha^2 m_{\tau}^2}
 \ \ Q^2 \ \left(1-\frac{Q^2}{m_{\tau}^2}\right)^2 
 \left(1+2\frac{Q^2}{m_{\tau}^2}\right) 
 \ \sigma(e^+e^-\to\pi^+\pi^-) 
 \labbel{CVCc}
\end{eqnarray}

\noindent
and the contributions to g-2 and \(\alpha(M_Z)\) are evaluated without
 kinematic corrections of phase space and form factor 
 the former
 are overestimated by 0.58\% and 0.16\%
 respectively (Tab.2).

 The situation is more complicated for the four pion case. All \(4\pi\)
 modes have different numbers of \(\pi^0\), whence the phase space 
 is different for each of the mode. Moreover, for a comparison between
 \(R(Q^2)\) and \(R^\tau(Q^2)\) one has to integrate over four particle
 phase space and it is impossible to obtain a simple analytical result.
  To estimate a size of the effect we 
  integrated the quantities \(R(--+0)\) etc. according to \Eq{diff}
 using 
 the current discussed in the next section once
 assuming that all masses are equal to \(m_-\) 
 and once taking the real masses. 
 For the mass corrections of the integrals we find
 
 \ba
 (-000)\ : \ 5.0  \% \ \ \ ; \ \ \ (--+0)\ : 2.4 \% \ \ \ ;\ \ \ 
 (++--)\ : \ 0 \% \ \ \ ; \ \ \ (+-00)\ : 4.6 \%
 \ea

 Mass effects alone thus modify the integrated version of the \Eq{diff} to

 \ba
 \frac{1}{1.050}\ \Gamma(-000) &=& \frac{1}{2} \ \Gamma(++--) \nonumber \\
 \frac{1}{1.024}\ \Gamma(--+0) &=& \frac{1}{2} \ 
  \Gamma(++--) + \frac{1}{1.046}\ \Gamma(+-00)
 \ ,
 \ea

\noindent
where 

\ba
 &&\Gamma(-000)=\Gamma(\tau^-\to \nu_\tau \pi^- 3\pi^0) \ \ ; \ \
      \Gamma(--+0)=\Gamma(\tau^-\to \nu_\tau 2\pi^- \pi^+ \pi^0) \nonumber \\
 &&\Gamma(++--)=  2 \ \Gamma_e \frac{\cos^2\theta_c}{m_{\tau}^2} \int
 \left(1-\frac{Q^2}{m_{\tau}^2}\right)^2 
 \left(1+2\frac{Q^2}{m_{\tau}^2}\right) R(++--) \ dQ^2 \nonumber \\
 &&\Gamma(+-00)=  2 \ \Gamma_e \frac{\cos^2\theta_c}{m_{\tau}^2} \int
 \left(1-\frac{Q^2}{m_{\tau}^2}\right)^2 
 \left(1+2\frac{Q^2}{m_{\tau}^2}\right) R(+-00) \ dQ^2 \ .
\ea

In principle this correction depends on the form of the current which
  will be specified in the next section.
  However, since \(a_1\) and \(\omega\) dominance
 and the qualitative behavior of the matrix element
 is well established
  this result will be applicable also to other forms of the current.
Additional effects could arise from modifications of the widths in 
 Breit-Wigner functions. The corrections are thus just indicative and
 detailed studies would be required which are beyond the scope of this work.

\begin{figure}[htbp]
\epsfig{figure=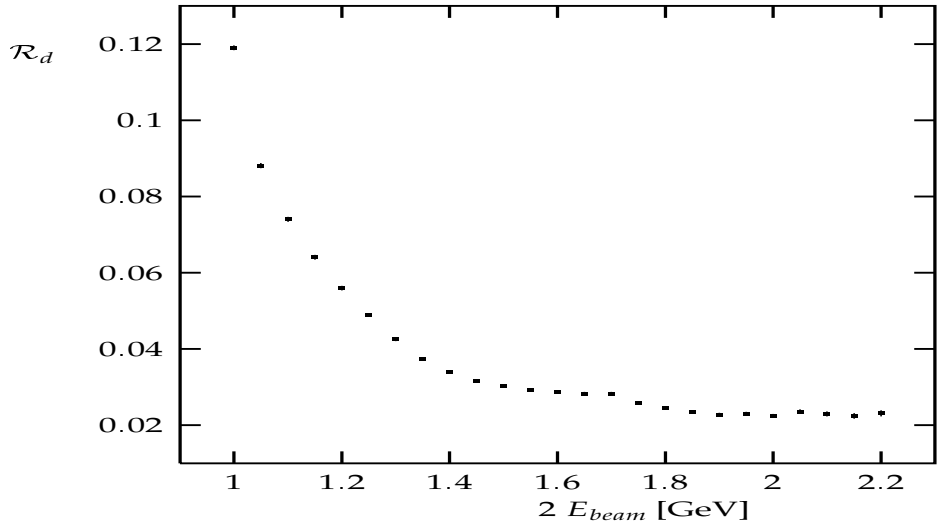, width=0.75\textwidth,height=6.5cm}
\caption{The ratio
 \( {\cal R}_d \equiv  \ \frac{\sigma_1-\sigma_2}{\sigma_2}\), where
 \(\sigma_1\) (\(\sigma_2\)) is \(e^+e^-\to 2\pi^0 \pi^-\pi^+\) cross section
 calculated for true pion masses (with all masses equal to \(m_-\)).
}
\label{f8}
\end{figure}

 It is also instructive to consider the mass effects on the differential
 rate for the \(e^+e^- \to 2\pi^0 \pi^+\pi^-\) process.
In Fig.\ref{f8} we plot the ratio 
 \({\cal R}_d \equiv \frac{\sigma_1-\sigma_2}{\sigma_2}\), where
 \(\sigma_1\) (\(\sigma_2\)) is the \(e^+e^-\to 2\pi^0 \pi^-\pi^+\)
  cross section
 calculated for true pion masses (with all masses equal to \(m_-\)).
 The difference amounts again
 up to a few percent.
The relations between the differential
  \(\tau\) decay rates and the \(e^+e^-\) cross sections obtained
  in the previous section will be violated at that level of accuracy even
  if  \Eq{rel} holds. However, to test these predictions experimentally
more precise measurements of the \(e^+e^-\to 4\pi\) cross section are required,
  where now a typical systematic error amounts to roughly 15 \%.

\section{The hadronic current. }

 As one can see from section 3 it is enough to construct only
 the hadronic current for the \((+-00)\) mode, while the
 other ones can be obtained using the relations \Eq{rel}.
 Its construction was
 based on  \cite{Fink} with some small changes allowing for 
 preserving the relations \Eq{rel}. The basic building block
 of the current contains a part built on the assumption of  \( a_1\) vector
 dominance plus an \(\omega\) exchange contribution.
 However only by adding an \(f_0\)
  contribution one
 can recover the proper chiral limit \cite{FWW}.
  The complete current can be written as
 a sum of these three contributions
 
\begin{eqnarray}
 \Gamma^{\mu}_{\rho^0 \rightarrow 2\pi^0 \pi^+ \pi^-} =
 \Gamma^{\mu}_{a_1} +\Gamma^{\mu}_{f_0}+\Gamma^{\mu}_{\omega} \ .
 \labbel{sum}
\end{eqnarray}

\begin{figure}[ht]
\begin{center}
\epsfig{file=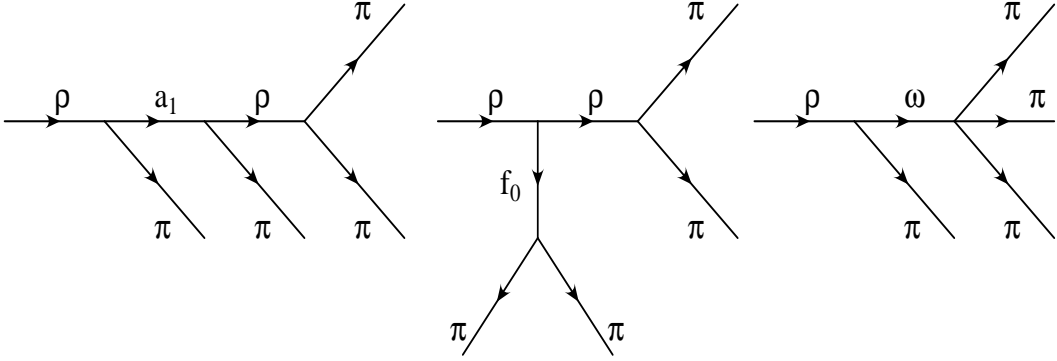,width=14cm,height=5cm}
\end{center}
\caption{Diagrams contributing to the hadronic current.
}
\label{f2}
\end{figure}

They are depicted schematically in Fig.\ref{f2} 
 and described in detail in the Appendix. Here we present
 some numerical tests of the current and a
 comparison between results obtained using the current \Eq{sum}
 and experimental data. One should add that the parameters of
 the model are kept as in \cite{Fink} even if in principle they
 should be re-fitted as the current is a bit different from the
 original one of \cite{Fink} and new and improved data have became
 available. This, however, is beyond the scope of this paper.  

 Starting from tests of the code of the current,
 first one can check if the Monte Carlo program
 reproduces the known \cite{tauola} analytical results
 of the partial \(\tau\) decay
 widths in the chiral limit

\ba
 \gamma_1 \equiv 
 \frac{\Gamma\left(\tau\to\nu_{\tau}2\pi^- \pi^+ \pi^0\right)}{\Gamma_e}
 = \frac{\cos^2\theta_c}{15} \left(\frac{m_\tau}{2\pi f_{\pi}}\right)^4
 \frac{1}{128}\left(\frac{1009}{96}-\pi^2\right) 
\labbel{width}
\ ,
\ea

\ba
 \gamma_3 \equiv 
 \frac{\Gamma\left(\tau\to\nu_{\tau}\pi^- 3\pi^0\right)}{\Gamma_e}
 = \frac{\cos^2\theta_c}{15} \left(\frac{m_\tau}{2\pi f_{\pi}}\right)^4
 \frac{1}{256}\left(\pi^2-\frac{437}{48}\right) \ .
\ea

\begin{table}
\begin{center}
$$
\begin{array}{|c|c|c|c|}
\hline
 {\rm Decay \  mode} 
& {\rm MC \ result \  (direct)}
& {\rm MC \ result \ (CVC)}
& {\rm analytical \  result}\\
\hline
 \gamma_1
&0.026786(4)
&0.026786(4)
&0.026788 \\
\hline
 \gamma_3
&0.015998(3)
&0.015996(2)
&0.015999 \\
\hline
\end{array}
$$
\vspace*{0.5cm}
\caption{
Comparison between analytical and Monte Carlo results in the chiral limit. 
 }
\end{center}
\label{t2}
\end{table}

 \Eq{width} differs from Eq.(38) of \cite{tauola}.
It was not discovered there that the analytical result is wrong as the tests
 were done at 0.2 \% (one sigma) precision level and the difference amounts
 to 0.34 \% . Tau decay rates can be obtained in two different ways. One way
 is just their direct calculation. The second one is by using the known
 relations to the \(e^+e^-\to 2\pi^+ 2\pi^- \) and 
 \(e^+e^-\to \pi^+\pi^- 2\pi^0\) cross sections \cite{Tsai,Fink}.
 The results of the numerical tests are summarized in 
 Tab.3, where CVC means the decay width was obtained through its relation
 to the \(e^+e^-\) cross sections.
As one can see the results of the test performed at 0.02 \% accuracy 
 level are quite satisfactory. 
 This agreement gives confidence in the numerical stability of our
 program.

\begin{table}
\begin{center}
$$
\begin{array}{|c|c|c|c|}
\hline
 {\rm Mode} 
& {\rm \cite{Fink}}
& {\rm present \  model}
& {\rm experiment}\\
\hline
 {\rm Br}\left( \tau^-\to\nu_{\tau}2\pi^-\pi^+\pi^0 \right)
&3.11\%
&4.33\%
&4.20(8)\% \\
\hline
 {\rm Br}\left( \tau^-\to\nu_{\tau}\pi^-
  \omega\left(\pi^-\pi^+\pi^0\right) \right) 
&1.20\%
&1.48\%
&1.73(11)\% \\
\hline
 {\rm Br}\left( \tau^-\to\nu_{\tau}\pi^-3\pi^0 \right)
&0.98\%
&1.14\%
&1.08(10)\% \\
\hline
\end{array}
$$
\vspace*{0.5cm}
\caption{
 Branching ratios of \(\tau\) decay modes. Results of \cite{Fink} 
 and the present current are compared to experimental data. 
 }
\end{center}
\label{t3}
\end{table}

Now we can test the physical predictions of the current. Let us start
 with \(\tau\) decay branching ratios. The results are
 summarized in Tab.4, where we put for completeness  also the results 
 presented in \cite{Fink} and the experimental results 
 \cite{PDG}. The agreement of the predictions of the current \Eq{sum}
 with the experimental data is satisfactory for 
 \(\tau^-\to\nu_{\tau}\pi^-3\pi^0\) decay mode.
Comparing however the results for the  
 \(\tau^-\to\nu_{\tau}2\pi^-\pi^+\pi^0\) and
  \(\tau^-\to\nu_{\tau}\pi^-
  \omega\left(\pi^-\pi^+\pi^0\right)\)
  modes
 it seems that the \(\omega\)
 part of the current does not represent the data well, even if
 the total branching ratio for the \(\tau^-\to\nu_{\tau}2\pi^-\pi^+\pi^0\)
 decay mode
 agrees with the data. Again the results of the partial
 decay widths were obtained as in the chiral limit by direct calculation
 and checked by relating them to the simulated \(e^+e^-\) cross sections.
 Agreement was found within statistical errors  proving that the
 code of the current fulfills the CVC relations \Eq{CVC}
 (if integrals are performed with \(m_-=m_0\)).

\begin{figure}[htbp]
\epsfig{figure=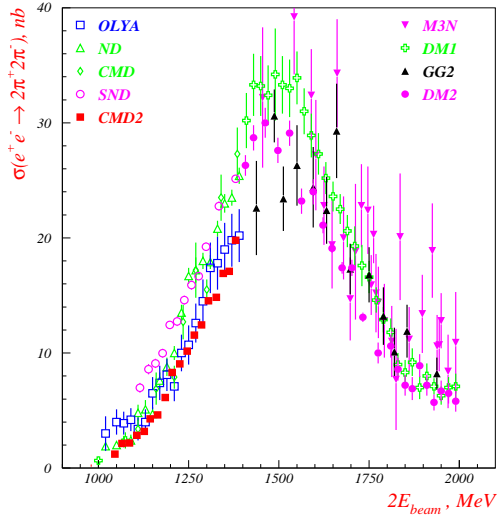, width=0.45\textwidth,height=7cm}
\hfill
\epsfig{figure=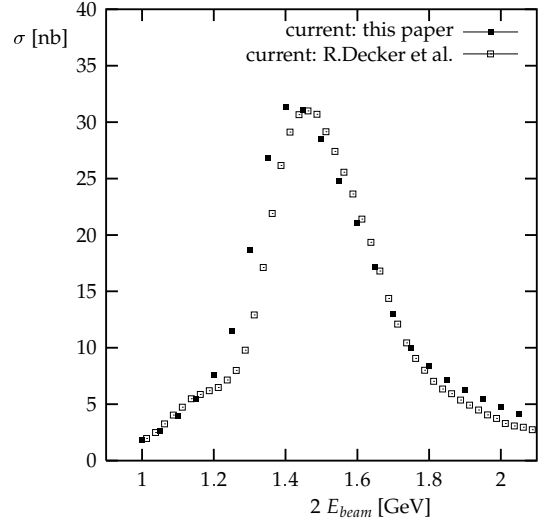,width=0.414\textwidth ,height=6.914cm}
\caption{
Comparison of data (left figure, see text) and predictions (right figure) 
obtained using the current \Eq{sum}
 (filled squares) and those of \cite{Fink} (empty squares) respectively
for $\sigma(e^+e^- \to 2 \pi^+2\pi^-)$.
}
\label{f3}
\end{figure}

\begin{figure}[htbp]
\epsfig{figure=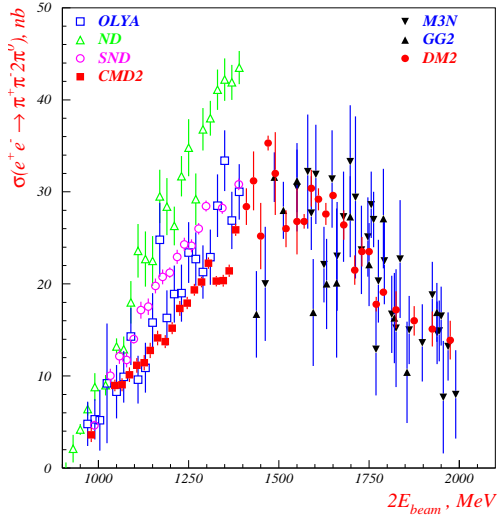, width=0.45\textwidth,height=7cm}
\hfill
\epsfig{figure=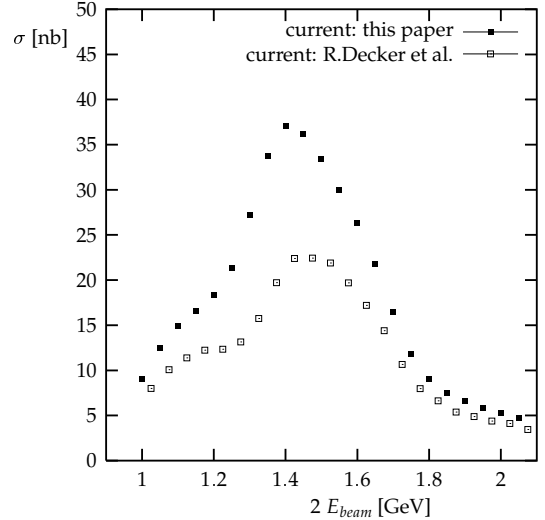,width=0.414\textwidth ,height=6.914cm}
\caption{
Comparison of data (left figure, see text) and predictions (right figure) 
obtained using the current \Eq{sum}
 (filled squares) and those of \cite{Fink} (empty squares) respectively
for $\sigma(e^+e^- \to \pi^+\pi^-2\pi^0 )$.
}
\label{f4}
\end{figure}

In the next step the predictions for the \(e^+e^-\to 2\pi^+ 2\pi^-\)
 (Fig.\ref{f3}) and \(e^+e^-\to \pi^+\pi^-2\pi^0\) (Fig.\ref{f4})
  cross sections
 are compared with data.
The plots taken from \cite{CMD2} 
contain data sets from OLYA \cite{OLYA}, ND \cite{ND1,ND2},
 CMD \cite{CMD}, SND \cite{SND}, CMD2 \cite{CMD2} plus results
 of the Orsay \cite{O1,O2,O3} and Frascati \cite{Fr1,Fr2,Fr3} groups.
 The data have a typical systematic error
 of about 15\% (shown in plots only for some of the data sets) and
 we have thus decided not to refit the parameters entering
 the current as the agreement between the Monte Carlo and the data
 is acceptable. The modification of the current of \cite{Fink}
 we performed leads to significantly better agreement between the theoretical
 predictions and the data in the mode \(e^+e^-\to \pi^+\pi^-2\pi^0\).
 Considering the agreement between the \(\tau\) decay rates and
 the prediction we
 conclude that the model
 reproduces the data well even if the description of the \(\omega\) part
 of the current is not completely satisfactory.

\section{The Monte Carlo program.\(^3\)  }

\footnotetext[3]{The program is available
 upon request from the authors.}

 The idea behind the structure of the Monte Carlo program
 is to allow for a simple addition of new final state modes into the program
 and for a simple replacement of the current(s) of the existing modes.
 The program thus exhibits a modular structure. For
 the generation of the four momenta of the mesons no sophisticated method
 of a variance reduction was applied. This slows down the generation,
 but could be accounted for if a faster Monte Carlo generator 
 would be required.
 It has, however, the advantage of being universal and no
 change of the variance reduction method is required
 with each modification of the hadronic
 current.
 The process to be simulated by the program in its final stage is
 \(e^+ e^- \rightarrow \gamma \ + \ hadrons \) 
 with an exclusive description of final states,
even if till now
 only  \(\pi^+ \pi^-\), \(2\pi^0 \pi^+ \pi^-\) and \(2\pi^+ 2\pi^-\)
 hadronic final states are implemented. The LL radiative QED
 corrections were taken into account using structure function method
 as developed in \cite{CCR} and limited to the initial emission only.
 In fact the program can run in one of two modes (chosen by a user)
  one with collinear radiation
 and one without it.
 Hard large angle photon 
 emission is limited to initial state radiation,
  which is justified by \cite{BKM} where it was demonstrated for the 
 \(\pi^+ \pi^-\) hadronic state that the contribution from the final
 state emission as well as the initial-final state interference can be
 reduced to a negligible level by applying suitable cuts.
 
 The generation of the multi particle phase space is based on the
 following representation of a Lorentz invariant phase space

 \begin{eqnarray}
 &&{\kern-15pt}{\rm dLips}_{n+1}\left(P,k,q_1,...,q_n\right)=
 \frac{1}{\left(2\pi\right)^{n-1}} \ \ {\rm dQ}_1^2 \ ... \ {\rm dQ}_{n-1}^2 
   \ {\rm dLips}_2\left(P,Q_1,k\right) 
 \ {\rm dLips}_2\left(Q_1,Q_2,q_1\right)
 \nonumber \\
 &&{\kern+120pt}\ \ ... \ \ {\rm dLips}_2\left(Q_{n-2},Q_{n-1},q_{n-2}\right)
 {\rm dLips}_2\left(Q_{n-1},q_{n-1},q_n\right) \ ,
 \end{eqnarray}  

\noindent
where \(P\) is a total four momentum of the \(photon \ + \ hadrons\) state
(it is not equal to the sum of the initial \(e^+ e^-\) four momenta as we
 allow for an additional initial collinear emission), \(k\) is the photon
 four momentum, \( q_1\ ... \ q_n\) are the four momenta of the hadrons,

 \begin{eqnarray}
 Q_1 = q_1 + \  ... \ +q_n \ \ {\rm and} \ \ Q_i = Q_{i-1}-q_{i-1} 
 \ \ {\rm for} \ \
 i=2,\ ... \ n-1 \ \ ,
 \end{eqnarray}  
 
 and \({\rm dLips}_2\left(k_1,k_2,k_3\right)\) is a two body phase space

 \begin{eqnarray}
 {\rm dLips}_2\left(k_1,k_2,k_3\right) = \frac{1}{32\pi^2} \ 
 \frac{\lambda^{\frac{1}{2}}\left(k_1^2,k_2^2,k_3^2\right)}{k_1^2} \ 
 {\rm d}\Omega_3
 \end{eqnarray}

 with   \(\lambda\left(a,b,c\right)= a^2+b^2+c^2-2ab-2ac-2bc\) and 
 \({\rm d}\Omega_3\) the \({\bf k}_3\) solid angle.
  
 The generation flow is the following:
first the collinear radiation is generated and a four momentum \(P\)
 of the visible final state is calculated and then boosted to its
 rest frame(RF) (a CM of the visible final state). This part is omitted
 in the mode running without collinear emission and then \(P\) is a sum
 of the electron and positron four momenta.
 In the second step the visible hard photon four momentum is generated.
 Its energy is generated flat even if the energy distribution is govern
 by a competition between soft \(\sim \frac{1}{E_\gamma}\) and a
 complicated resonant spectrum which depends on details of the hadronic
 current. Of course this way the generator is not extremely efficient,
 but it is more universal and will work equally well with all reasonable
 modifications of the hadronic current. The photon polar angle is
 generated in \(P\) RF with the distribution accounting 
 for collinear emission peaks,
 which are the same for all hadronic modes 
 as the initial state is ever the same.
 Its azimuthal angle again is generated with a flat distribution in \(P\) RF.
   Then a chain of a generations
 of \(Q_i^2\ , \ i=2,...,n-1\) follows (\(Q_1\) is fixed when we generate
 the photon four momentum). 
 They are generated flat within their allowed limits

 \begin{eqnarray}
  \left(\sum_{k=i}^{n}m_k\right)^2\ < \ Q_i^2 \ 
  < \ \left(Q_{i-1}^{(0)}-m_{i-1}\right)^2
  \end{eqnarray}

where \(Q_i^{(0)}\) is a zeroth component of the \(Q_i\) four momentum.
At the end a generation of the solid angles \(\Omega_i,\ i=1,...,n-1\)
 follows. They are generated flat ( \( \Omega_i \) in \(Q_i\) RF).
 All generated four momenta are then transformed with the use of the proper
 boosts and rotations into the CM system of the initial \(e^+e^-\) particles. 
 The distribution of the events \(\sim |{\cal M}|^2\), where \({\cal M}\) is
 a matrix element of a given process is obtained by means of the 
 rejection method. The cross section is calculated using \Eq{SIG}
 both for weighted and unweighted event samples. The unweighted events
 are stored in a file when requested.

 Let us now discuss the cuts, which
 reduce the contribution of the final state emission to the cross section 
 to a negligible level allowing thus for extraction of \(R(Q^2)\) from
  the measurement of the
 \(e^+e^-\to \ hadrons \ + \gamma\) cross section.
 We recall that in \cite{BKM} it was shown that the following set of 
 angular cuts 
 
\ba
 {\rm cuts1:} \ \ \ \ \ \ \ \ \ \
 \left( 7^{\circ} < \theta_\gamma < 20^{\circ} 
 \ \ \ \ \ or \ \ \ \ \ 160^{\circ} < \theta_\gamma < 173^{\circ}\right) 
 \ \ \  and 
  \ \ \ 30^{\circ} < \theta_\pi < 150^{\circ} \ ,
 \labbel{cuts}
\ea

\noindent
 ( \(\theta_\gamma\) (\(\theta_\pi\) ) is the photon (pion) polar angle)
 fulfils this requirement for the 
 \(e^+ e^- \to \pi^+ \pi^- \gamma\) cross section. It reduces, however,
 the observed cross section significantly. This starts to become dramatic, when
 one runs at energies well above 1 GeV. The following
 set of cuts

\ba
 {\rm cuts2:} \ \ \ \ \ \ \ \ \ \
 &&\left( 7^{\circ} < \theta_\gamma < 20^{\circ}  \ \ \
 and  \ \ \ 30^{\circ} < \theta_\pi < 173^{\circ}\right) \nonumber \\
   &&{\kern-40pt}or  \ \ \ 
 \left( 160^{\circ} < \theta_\gamma < 173^{\circ}  \ \ \
 and  \ \ \ 7^{\circ} < \theta_\pi < 150^{\circ}\right) \ ,
\labbel{cuts1}
\ea

\noindent
also reduces the contribution from final state radiation
 to a
 negligible level due to the fact that
 the pions and photon are well separated as in the
 previous case.

\begin{figure}[htbp]
\epsfig{figure=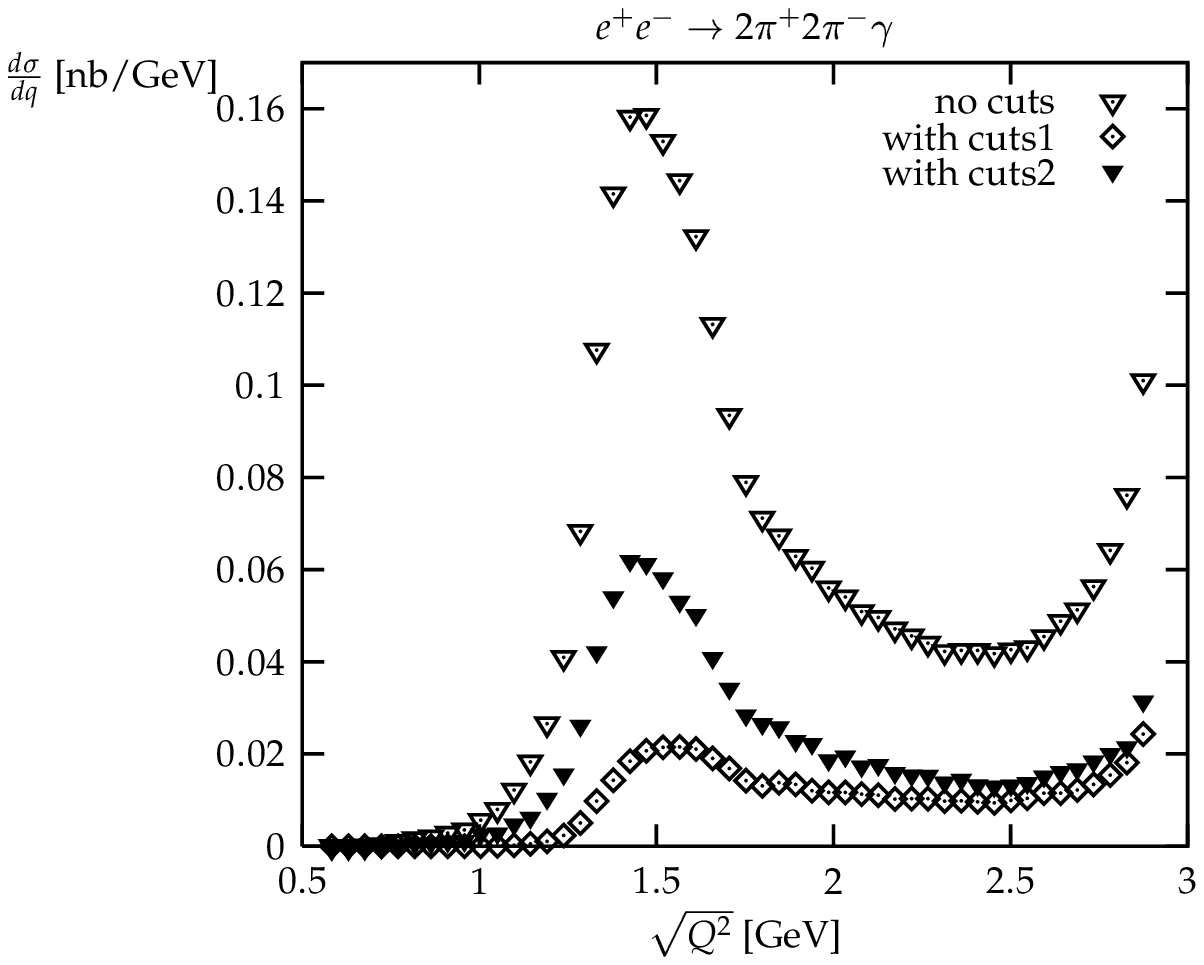, width=0.45\textwidth,height=7cm}
\hfill
\epsfig{figure=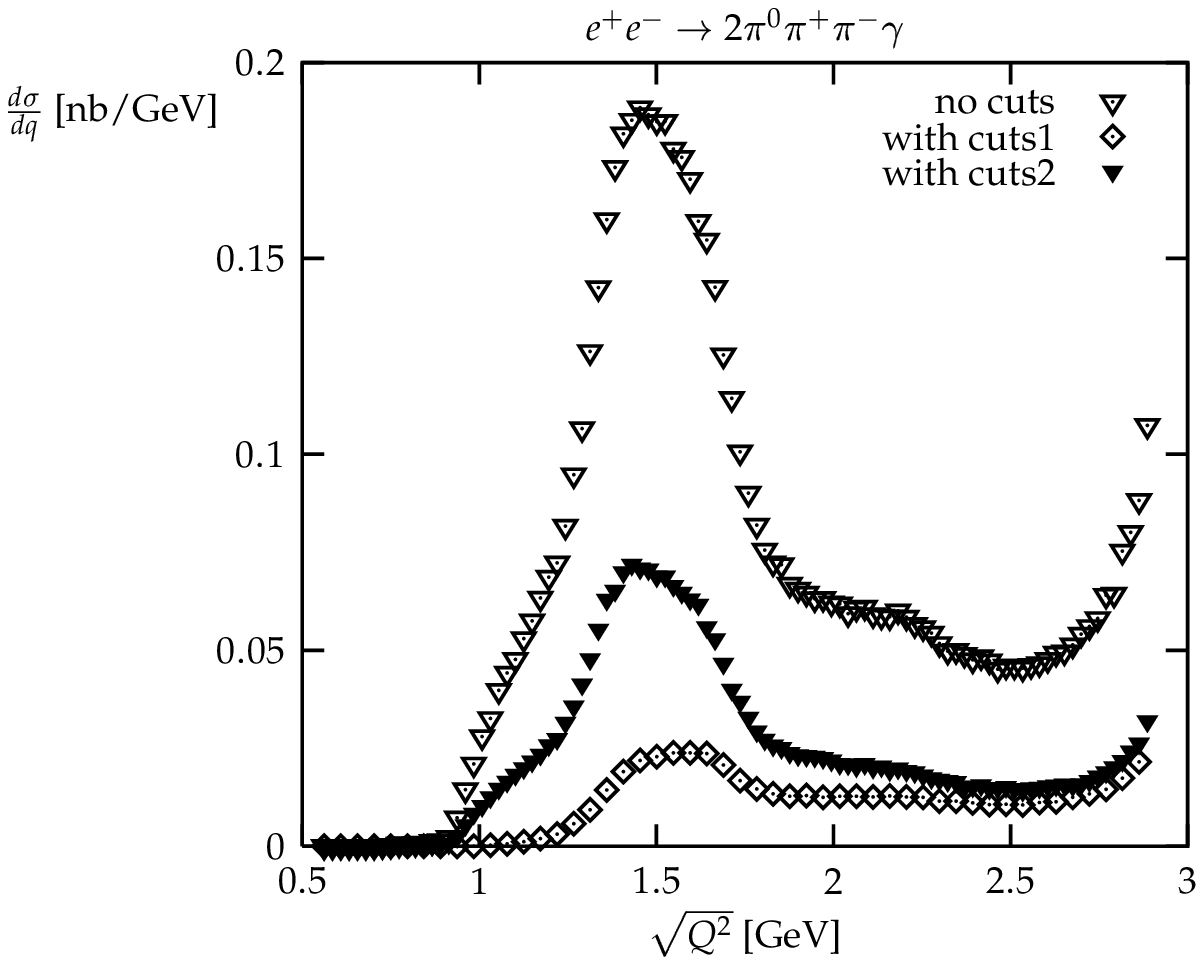,width=0.45\textwidth ,height=7cm}
\caption{The differential \(e^+e^-\to 4\pi\gamma\) cross sections at
 beam energy 1.5 GeV with minimal photon energy equal 0.1 GeV with
  no cuts on pions angles and \(7^{\circ} < \theta_\gamma < 173^{\circ}\)
 ( no cuts) and two sets of angular cuts \Eq{cuts}
 (with cuts1) and \Eq{cuts1} (with cuts2), where \(q\equiv \sqrt{Q^2}\)
 is an invariant mass of the \(4\pi\) system.
}
\label{f5}
\end{figure}

\begin{figure}[htbp]
\epsfig{figure=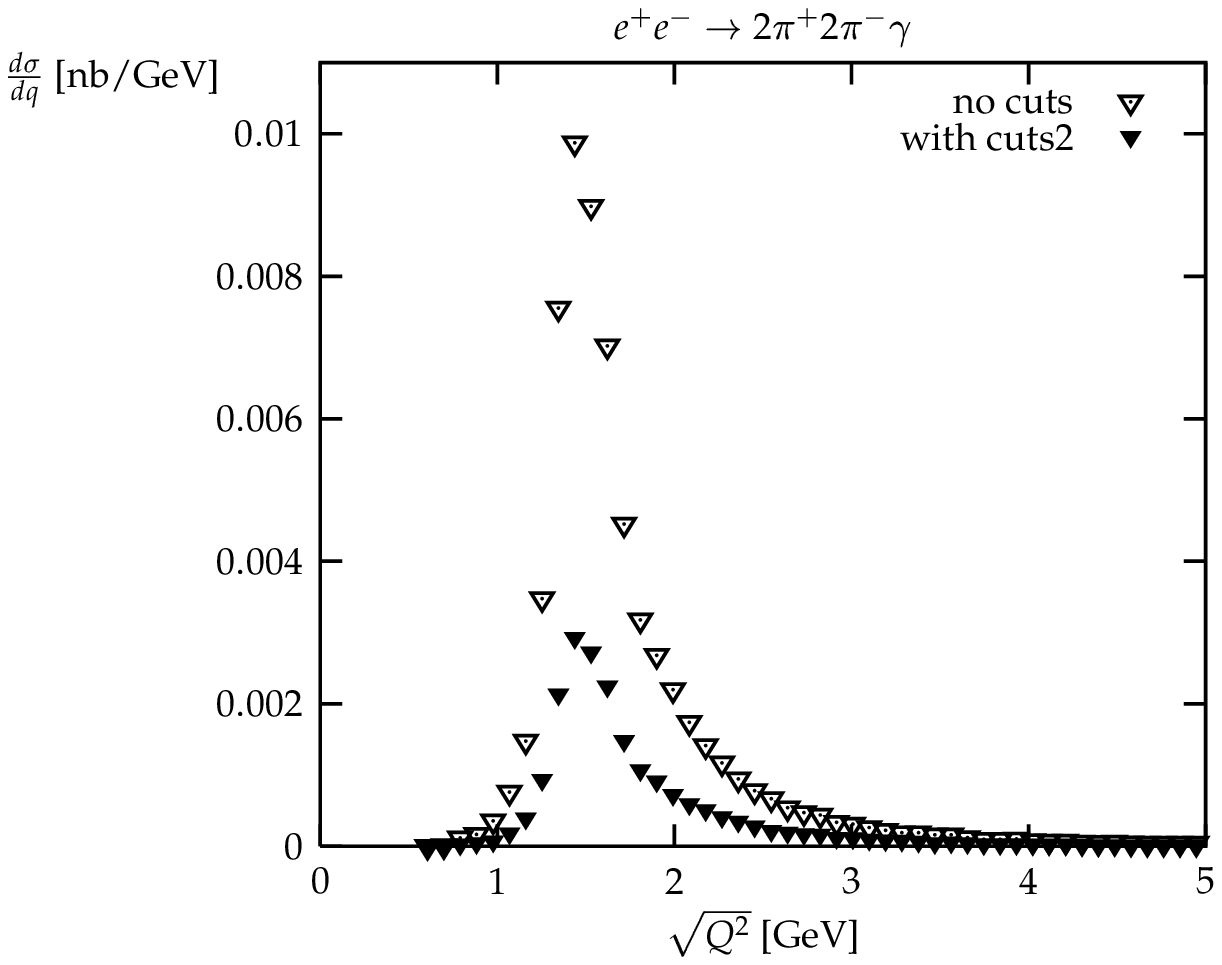, width=0.45\textwidth,height=7cm}
\hfill
\epsfig{figure=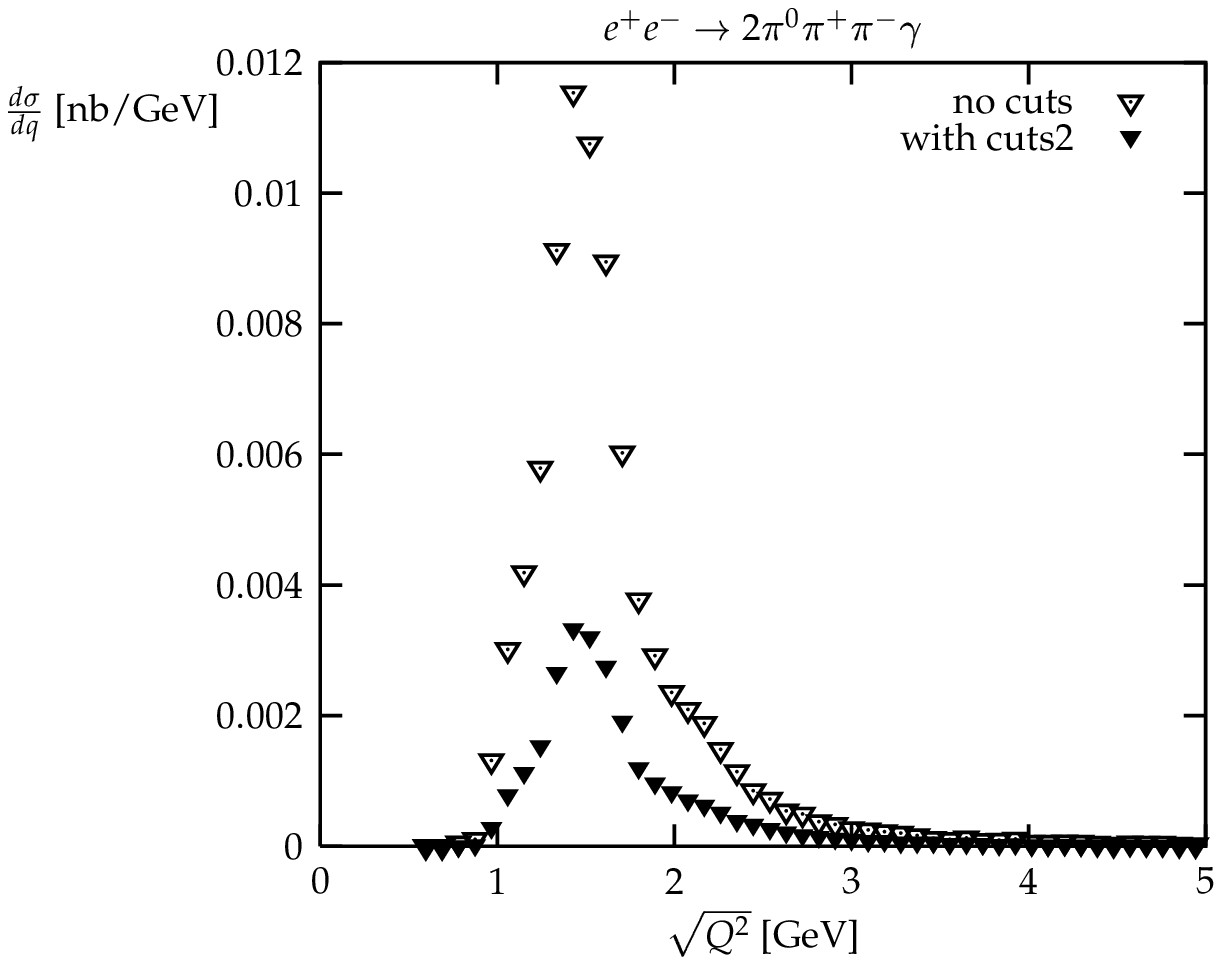,width=0.45\textwidth ,height=7cm}
\caption{The differential \(e^+e^-\to 4\pi\gamma\) cross sections at
 beam energy 5 GeV with minimal photon energy equal 0.2 GeV with
no cuts on pions angles 
 and \(7^{\circ} < \theta_\gamma < 173^{\circ}\)( no cuts) 
 and angular cuts of \Eq{cuts1}
 (with cuts2), where \(q\equiv \sqrt{Q^2}\)
 is an invariant mass of the \(4\pi\) system.
}
\label{f6}
\end{figure}

At the same time the cross section reduction is much
 smaller, especially for higher beam energies and higher energies
 of the observed photons. The effect of the two sets of cuts on the
 cross sections with four pions in the final state is presented 
in Fig.\ref{f5} for  \(2E_{beam}\) = 3 GeV. 
 For higher beam energies the effect of the cross section
 reduction is much higher and at 2 \(E_{beam}\) = 10 GeV the cross
 sections for cuts specified in \Eq{cuts} is reduced almost to zero. 
 For the cuts specified in \Eq{cuts1} the results
 are presented in Fig.\ref{f6} and the reduction remains tolerable.

\begin{table}
\begin{center}
$$
\begin{array}{||c||c||c|c||}
\hline
&
&
\multicolumn{2}{|c|}{{\rm Event~rates}}\\
\hline
\sqrt{s}
&{\rm Integrated~luminosity}, {\rm fb}^{-1} 
& \ \ \ 2\pi^+2\pi^-\gamma \ \ \ 
& \ \ \ 2\pi^0\pi^+\pi^-\gamma \ \ \ 
\\ \hline \hline
\ 1 \ {\rm GeV}\ & 1 & 1.04\cdot 10^4 & 1.13 \cdot 10^4 
 \\ \hline \hline
\  3\ {\rm GeV}\ & 1 & 4.66 \cdot 10^4 & 5.72 \cdot 10^4 
  \\ \hline \hline
\  10\ {\rm GeV}\  & 100 & 1.86 \cdot 10^5 & 2.33 \cdot 10^5 \\ 
\hline
\end{array}
$$
\caption{Estimated number of  radiative 
events $e^+e^- \to \ 4\pi \ + \ \gamma$ for different center of mass
energies. The minimal photon energy is: 0.05 GeV (first row),
 0.1 GeV (second row), 0.2 GeV (third row). The angular cuts of \Eq{cuts1}
 were applied.
 }
\end{center}
\label{t4}
\end{table}

From Fig.\ref{f6} one concludes that one can measure
 \(R(Q^2)\) at a B factory in the interesting
  region of \( \sqrt{Q^2}\) between 1 GeV and 2.5 GeV  through
  measuring the \(e^+e^-\to 4\pi\gamma\) cross section
 using \Eq{1}. This measurement should have an an accuracy
 much better then 15 \% , which is now the typical
 experimental error in that \(Q^2\) region, 
 allowing for a reduction of the error
 in the calculation of the photon vacuum polarization. 
 From  Table 5 it is clear that the 
  error would be dominated by systematics and not by statistics.

One may even restrict photon and pions detection angles to the central
 region, e.g. 
 \( 25^{\circ} < \theta_\gamma < 155^{\circ}\)
 and   \( 30^{\circ} < \theta_\pi < 150^{\circ}\) respectively
 if a minimal angle of \(20^{\circ}\) between photon and charged 
 and neutral pions is required in order to suppress final state
 radiation and to clearly separate neutral pions and the photon.
 With this cut one obtains (\(\sqrt{s}= 10 \ GeV\) and \({\cal L}
 = 100 \ fb^{-1} \) ) a rate of \(1.17\cdot 10^5\) events with
 \(2\pi^+ 2\pi^- \gamma\) and \(1.53 \cdot 10^5\) events with
 \(2\pi^0\pi^{+}\pi^{-}\gamma\).

\begin{figure}[htbp]
\epsfig{figure=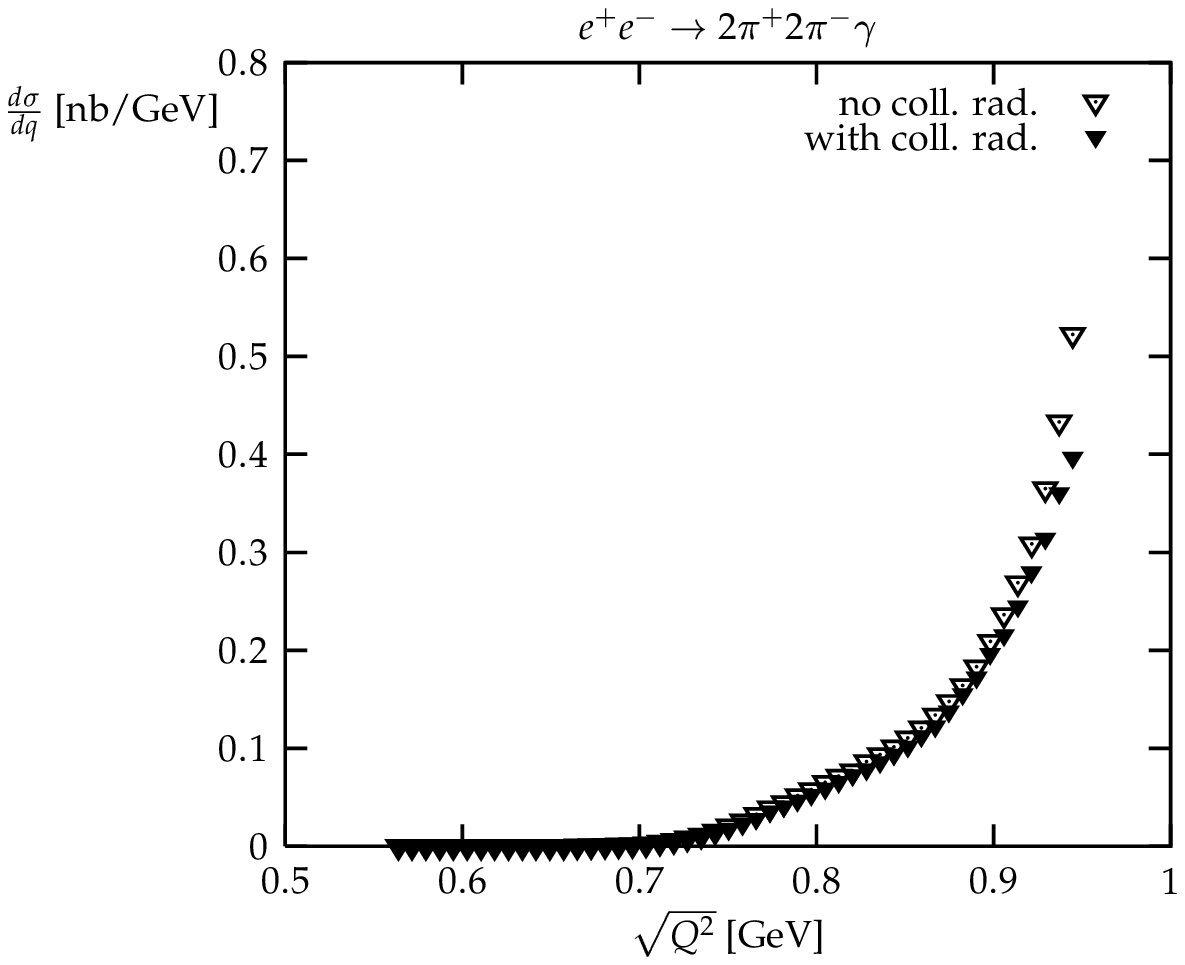, width=0.45\textwidth,height=6.5cm}
\hfill
\epsfig{figure=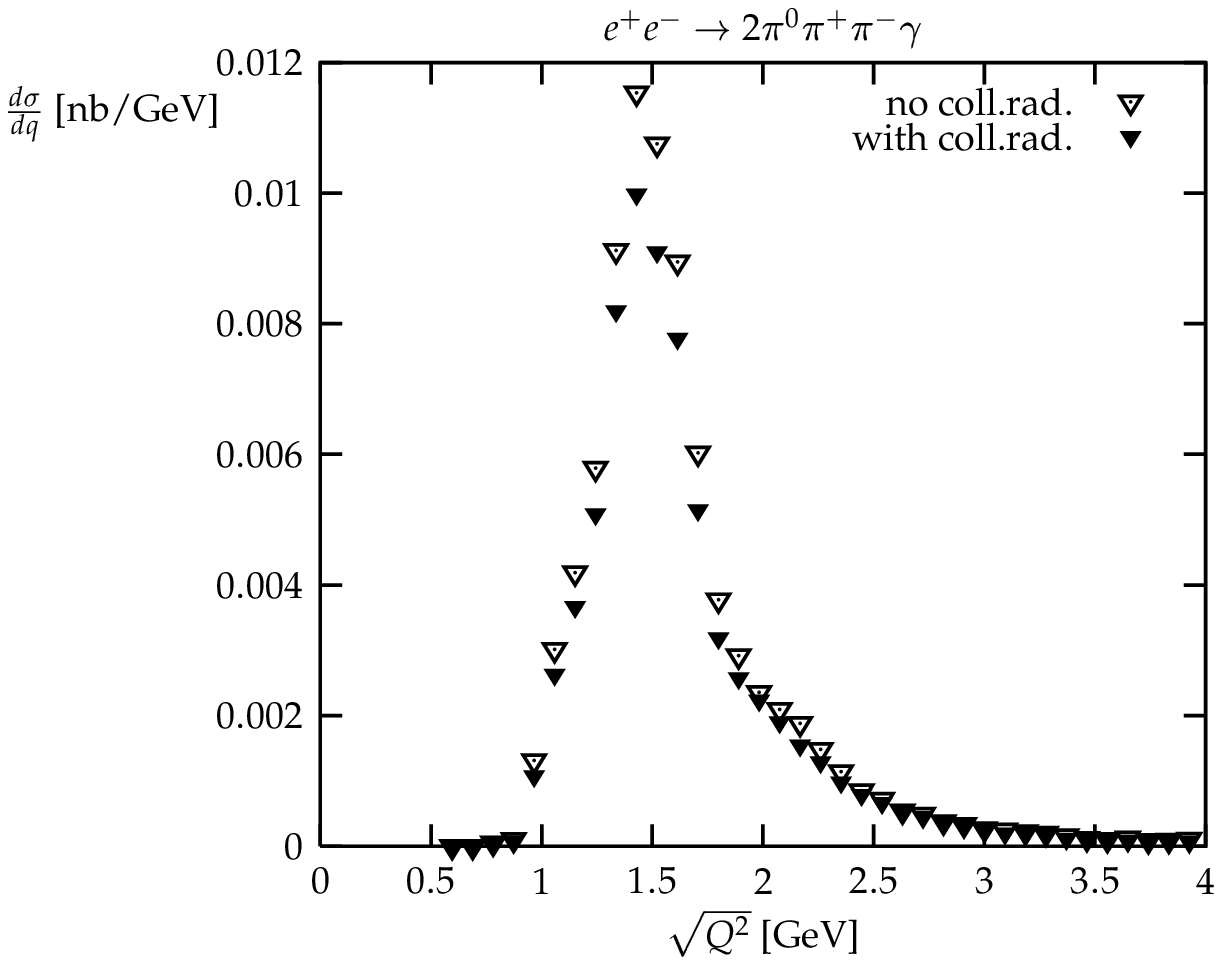,width=0.45\textwidth ,height=6.5cm}
\caption{The effect of the collinear radiation on the 
 differential \(e^+e^-\to 4\pi\gamma\) cross sections at
 beam energy 0.5 GeV (left picture) and 5 GeV (right picture)
  with minimal photon energy equal to 0.05 and 0.2 GeV 
 correspondingly. In the mode with collinear radiation minimal
 invariant masses of the \(4\pi \ + \ \gamma\) systems  
 of 0.95 GeV and 9.5 GeV were required. 
  \(q\equiv \sqrt{Q^2}\)
 denotes the invariant mass of the \(4\pi\) system.
}
\label{f7}
\end{figure}

 Additional collinear emission always present in the real experiment
 reduces slightly the cross sections as shown in Figs.\ref{f7} for
 two different modes and two different beam energies.
 Its actual size depends on the cuts on the invariant mass of
 the \(4\pi\ +\ \gamma\) system. The effect is similar for different
 energies and for both charge modes.

\section{Summary.}
A precise value of the cross section for hadron production in electron
positron annihilation at low energies is one of the important
ingredients for a reliable prediction of the anomalous magnetic of the
muon and the electromagnetic coupling at high energies. As an alternative
to a direct measurement at the relevant energy one may use initial state
radiation to reduce the effective energy of electron positron colliders, 
exploiting the large luminosity of ``factories'' and accessing thus a
continuum of hadronic final states.

With this motivation a Monte Carlo generator has been constructed to
simulate the reaction $e^+e^- \to \gamma + 4 \pi$, where the photon is
assumed to be observed in the detector. The hadronic matrix element has
been taken from \cite{Fink}. 
Isospin relations between the amplitudes
governing $\tau$ decays into four pions and electron positron
annihilation into four pions have been found which allow to
determine all four modes after the amplitude for the $\pi^+\pi^-2\pi^0$
channel  has been fixed. The kinematic breaking of these isospin
relations as a consequence of the $\pi^-$ -- $\pi^0$ mass difference has
also been investigated.

The program is constructed in analogy to
the one \cite{BKM} simulating $e^+e^- \to \gamma  + 2 \pi$. However, it does
not include final state radiation from the charged pions. Additional
collinear photon radiation has been incorporated with the technique of
structure functions. Predictions are presented for cms energies of 1GeV,
3GeV and 10GeV, corresponding to the energies of DA\(\Phi\)NE, BEBC and
of $B$-meson factories.

Even after applying realistic cuts the event rates are sufficiently high
to allow for a precise measurement of $R(Q^2)$ in the region of $Q$
between approximately 1GeV and 2.5GeV.

The model predictions are compared to recent data from electron positron
colliders. Once more accurate data become available, the modular
structure of the program will allow for modification of
replacement of the hadronic current in a simple way.


\vskip 0.4 cm

{\bf Acknowledgments.}
The work by H.C. was supported by BMBF-POL-239-96, 
the work by J.K. by BMBF under grant BMBF-057KA92P. 
H.C. is grateful for the support and the kind hospitality 
of the Institut f{\"u}r Theoretische Teilchenphysik
 of the Karlsruhe University, where this work was carried out. 

\vskip 0.4 cm

\begin{appendix}

\noindent
{\Large \bf Appendix: The description of the hadronic current.}

 In this appendix we give a complete definition of the hadronic current
 used in this paper. Recalling \Eq{sum}
  
\begin{eqnarray}
 \Gamma^{\mu}_{\rho^0 \rightarrow 2\pi^0 \pi^+ \pi^-} =
 \Gamma^{\mu}_{a_1} +\Gamma^{\mu}_{f_0}+\Gamma^{\mu}_{\omega} \ ,
\end{eqnarray}

\noindent
 we will define here its ingredients.

 Denoting  the pions four momenta
 as follows: \(q_1(\pi^0) ,q_2(\pi^0) , q_3(\pi^-) \ \ {\rm and}\ 
 \ q_4(\pi^+) \)
 one
 gets the following contribution from the part containing an \(a_1\) exchange

\begin{eqnarray}
 &&\Gamma^{\mu}_{a_1}\left(q_1,q_2,q_3,q_4\right) = \nonumber \\ 
 &&\tilde \Gamma^{\mu}_{a_1}(q_3,q_2,q_1,q_4) +
 \tilde \Gamma^{\mu}_{a_1}(q_3,q_1,q_2,q_4) -
 \tilde \Gamma^{\mu}_{a_1}(q_4,q_2,q_1,q_3) -
 \tilde \Gamma^{\mu}_{a_1}(q_4,q_1,q_2,q_3). 
 \end{eqnarray}

The structure of this current follows from the form of the Lagrangian
of the \(\rho a \pi \) (\( \sim \vec a \cdot \left(\vec\pi \times 
 \vec \rho\right)\) ) and \(\rho \pi \pi \) 
 (\( \sim \vec \rho \cdot \left(\vec\pi_1 \times 
 \vec \pi_2\right)\) ) interactions and the function 
 \(\tilde \Gamma^{\mu}_{a}\) is defined as \cite{Fink}

\begin{eqnarray}
 \tilde \Gamma^{\mu}_{a_1}(q_1,q_2,q_3,q_4) &=& 
  C_1 \ \
 T_{\rho}\left(Q^2\right)  \ \ \Gamma^{\mu \nu}\left(Q,Q-q_1\right)
 \nonumber \\
  &&BW_{a_1}\left((Q-q_1)^2\right) \ \
 \Gamma_{\nu\lambda }\left(Q-q_1,Q-q_1-q_2\right)\nonumber \\
 &&BW_{\rho}\left((q_3+q_4)^2\right) \ \
 \Gamma_1^{\lambda}(q_3-q_4) \ ,
 \end{eqnarray} 

 where we have followed the notation of \cite{Fink}; 
 \( Q = q_1+q_2+q_3+q_4\).

 Assuming that \(q_1^2 = q_2^2=q_3^2=q_4^2=m_{\pi}^2 \) i.e. that
 pions have equal masses one finds

\begin{eqnarray}
 &&{\kern-30pt}\tilde \Gamma^{\mu}_{a_1}(q_1,q_2,q_3,q_4) = \nonumber \\
  &&C_1 \  f_{\rho\pi\pi}\left((q_3+q_4)^2\right) \ 
          f_{a_1\rho\pi}(Q^2,(Q-q_1)^2) \ f_{a_1\rho\pi}((Q-q_1)^2,(q_3+q_4)^2)
\nonumber \\
 &&T_{\rho}\left(Q^2\right)  \ \ BW_{a_1}\left((Q-q_1)^2\right) \ \
 BW_{\rho}\left((q_3+q_4)^2\right) \nonumber \\
 &&{\kern-7pt} 
 \left[(q_3-q_4)^{\mu} + q_1^{\mu} \frac{q_2(q_3-q_4)}{(Q-q_1)^2}
 -Q^{\mu}\left(\frac{(q_1+q_2)(q_3-q_4)}{Q^2} 
 +\frac{(Q q_1)(q_2(q_3-q_4))}{Q^2 (Q-q_1)^2}\right)\right] \ ,
\end{eqnarray}

where \(f_{\rho\pi\pi}\) and \(f_{a_1\rho\pi}\) describe \(\rho \pi \pi\)
and \(a_1\rho\pi\) vertices appropriately while \(T_{\rho}\),
 \(BW_{a_1}\) and \(BW_{\rho}\) are \(\rho\), \(a_1\) and again \(\rho\)
 propagators. The different choice of the authors of \cite{Fink}
 for two \(\rho\) propagators comes from the fact that \(\rho\)
 couples in a different way in the two cases and the 'propagators' absorb
 in fact some parts of the couplings. The normalization constant 
 \(C_1 =\frac{2 \ \sqrt{6}}{f_{\pi}^2}\) was chosen to give 
 the proper chiral limit of the current. The form factors were chosen
 to be constant 

 \( \left[f_{\rho\pi\pi}\left((q_3+q_4)^2\right) \ 
 f_{a_1\rho\pi}(Q^2,(Q-q_1)^2) \ f_{a_1\rho\pi}((Q-q_1)^2,(q_3+q_4)^2)
 \right]= 0.38\),
 
\noindent
 and the validity of that assumption was proved \cite{Fink} 
 in a limited range of \(Q^2\)  

\noindent
 (\(1.1 \ {\rm GeV} \ < \sqrt{Q^2} \ < \ 2.2 \ {\rm GeV} \) ).
   
 The contribution to the current from \(f_0\) exchange reads

\begin{eqnarray}
 \Gamma^{\mu}_{f_0}(q_1,q_2,q_3,q_4)  &=& \nonumber \\
 &&{\kern-20pt}C_2 \ f_{\rho \rho f_0}\left(Q^2,(q_3+q_4)^2,(q_1+q_2)^2\right) 
 \ f_{\rho\pi\pi}\left((q_3+q_4)^2\right)
                  \ f_{f_0 \pi\pi}\left((q_1+q_2)^2\right)\nonumber \\
 &&T_{\rho}\left(Q^2\right) \ T_{\rho}\left((q_3+q_4)^2\right) \ 
  BW_{f_0}\left((q_1+q_2)^2\right)
\nonumber \\
 &&\left[(q_3-q_4)^{\mu} - Q^{\mu} \frac{Q(q_3-q_4)}{Q^2}\right] \ ,
\end{eqnarray}

 where \(f_{\rho \rho f_0}\) and \( f_{f_0 \pi\pi}\) 
 describe \(\rho \rho f_0 \)
and \( f_0 \pi\pi\) vertices appropriately, while \(BW_{f_0}\) is an 
 \( f_0 \) propagator. 

Again \( C_2 = -\frac{3 \ \sqrt{6}}{f_{\pi}^2}\)
 follows from the chiral limit and the form factors are kept constant
 
\( f_{\rho \rho f_0}\left(Q^2,(q_3+q_4)^2,(q_1+q_2)^2\right) 
 \ f_{\rho\pi\pi}\left((q_3+q_4)^2\right)
                  \ f_{f_0 \pi\pi}\left((q_1+q_2)^2\right) = 0.38 \ . \)

 The contribution coming from so called anomalous (containing \(\omega\)
 exchange) part of the current reads

\begin{eqnarray}
 &&\Gamma^{\mu}_{\omega}(q_1,q_2,q_3,q_4)=
 \frac{g_{\omega}} {\sqrt{2}} \ 1475.98 \ {\rm GeV}^{-3} 
 \ 12.924 \ {\rm GeV}^{-1} \ 0.266 \ m_{\rho}^2 \nonumber \\
 &&{\kern-25pt}
 \left[q_{1}^{\mu} F_1(q_1,q_2,q_3,q_4)+q_{2}^{\mu} F_1(q_2,q_1,q_3,q_4)
  +q_{3}^{\mu} F_2(q_1,q_2,q_3,q_4)-q_{4}^{\mu} F_2(q_1,q_2,q_4,q_3)
  \right] \ ,
 \labbel{om1}
\end{eqnarray}

where 
\begin{eqnarray}
&&{\kern-25pt}F_1(q_1,q_2,q_3,q_4)= BW_{\rho,\omega}(Q^2,(Q-q_2)^2)
 \left[(q_3(Q-q_2)) \ (q_2q_4) -(q_4(Q-q_2)) \ (q_2q_3) \right] \\
&&{\rm and}\nonumber \\
&&{\kern-25pt}F_2(q_1,q_2,q_3,q_4)= BW_{\rho,\omega}(Q^2,(Q-q_2)^2)
 \left[(q_4(Q-q_2)) \ (q_1q_2) -(q_1(Q-q_2)) \ (q_2q_4) \right]\nonumber \\
&&{\kern+55pt}+BW_{\rho,\omega}(Q^2,(Q-q_1)^2)
 \left[(q_4(Q-q_1)) \ (q_1q_2) -(q_2(Q-q_1)) \ (q_1q_4) \right] 
 \labbel{om2}
\end{eqnarray}

\noindent
and \(g_{\omega} = 1.55\). It is changed from its original value
 \(g_{\omega} = 1.4\) in \cite{Fink} to reproduce the \(\tau\)
 decay rates \(\tau^-\to\nu_{\tau}2\pi^-\pi^+\pi^0\) and 
 \( \tau^-\to\nu_{\tau}\pi^-\omega\left(\pi^-\pi^+\pi^0\right)\).

For completeness we list here all propagators \cite{Fink} required
for the current:

\begin{eqnarray}
T_{\rho}\left(Q^2\right)= \frac{
  BW_3(Q^2,m_{\rho},\Gamma_{\rho}) 
 + \beta_1 BW_3(Q^2,m_{\rho_1},\Gamma_{\rho_1})
 + \beta_2 BW_3(Q^2,m_{\rho_2},\Gamma_{\rho_2})}{1+\beta_1+\beta_2} \ ,
\end{eqnarray}

with

\begin{eqnarray}
 BW_3(Q^2,m_{\rho},\Gamma_{\rho}) = \frac{m_{\rho}^2}
 {m_{\rho}^2-Q^2 - i
 \Gamma_{\rho}m_{\rho}\sqrt{\frac{m_{\rho}^2}{Q^2} 
 \left[\frac{Q^2-4m_{\pi}^2}{m_{\rho}^2-4m_{\pi}^2}\right]^3} }
\end{eqnarray}

and the numerical values set to
\begin{eqnarray}
 && m_{\pi} = 0.14 \ {\rm GeV}\nonumber \\
 && m_{\rho} = 0.773 \ {\rm GeV} \ \ \Gamma_{\rho} = 0.145 
 \ {\rm GeV}\nonumber \\
 && m_{\rho_1} = 1.35 \ {\rm GeV} \ \ \Gamma_{\rho_1} = 0.3 
 \ {\rm GeV} \ \ \beta_1 = 0.08 \nonumber \\
 && m_{\rho_2} = 1.7 \ {\rm GeV} \ \ \Gamma_{\rho_2} = 0.235 
 \ {\rm GeV} \ \ \beta_2 = -0.0075  \ \ ;
\end{eqnarray}

\begin{eqnarray}
BW_{\rho}\left(Q^2\right) = \frac{
  BW_3(Q^2,m_{\rho},\Gamma_{\rho}) 
 + \beta BW_3(Q^2,\tilde m_{\rho_1},\tilde \Gamma_{\rho_1})
  }{1+\beta} \ \ ,
\end{eqnarray}

with 
\begin{eqnarray}
\tilde m_{\rho_1}= 1.37 \ {\rm GeV} \ \ \tilde \Gamma_{\rho_1} = 0.510
 \ {\rm GeV} \ \ \beta = -0.145 \ ;
\end{eqnarray}

\begin{eqnarray}
 BW_{a_1}\left(Q^2\right) = \frac{m_{a_1}^2}
 {m_{a_1}^2-Q^2 - i \Gamma_{a_1}m_{a_1}\frac{g(Q^2)}{g(m_{a_1}^2)}} \ ,
\end{eqnarray}

where
\begin{eqnarray}
 &&g(Q^2) =  1.623 \ Q^2 + 10.38 - \frac{9.32}{Q^2} + \frac{0.65}
        {(Q^2)^2} \ \ \ \ {\rm for}  \ \ \ \ Q^2 > (m_{a_1}+m_{\pi})^2
  \nonumber \\
 &&g(Q^2) =  4.1 \ \left(Q^2-9m_{\pi}^2\right)^3
 \left[ 1 -3.3 \left(Q^2-9m_{\pi}^2\right) + 5.8\left(Q^2-9m_{\pi}^2\right)^2
 \right] \nonumber \\
  &&{\kern+100pt}{\rm for}  \ \ Q^2 < (m_{a_1}+m_{\pi})^2
\end{eqnarray}

and 
\begin{eqnarray}
 m_{a_1} = 1.251 \ {\rm GeV} \ \  \Gamma_{a_1} = 0.599 \ {\rm GeV} \ ;
\end{eqnarray}

\begin{eqnarray}
BW_{f_0}\left(Q^2\right) =  \frac{m_{f_0}^2 - i m_{f_0}\Gamma_{f_0}}
 {m_{f_0}^2 - Q^2 - i m_{f_0}\Gamma_{f_0}}
\end{eqnarray}

with 
\begin{eqnarray}
m_{f_0} = 1.3 \ {\rm GeV} \ \  \Gamma_{f_0} = 0.6 \ {\rm GeV} \ ;
\end{eqnarray}

\begin{eqnarray}
BW_{\rho,\omega}(Q^2,Q_1^2) &=& 
 \left[ \frac{1}{m_{\rho}^2 - Q^2 - i m_{\rho}\Gamma_{\rho}}
 +\sigma \frac{1}{m_{ \rho\prime}^2 - Q^2 - i m_{\rho\prime}
 \Gamma_{\rho\prime}}
 \right] \frac{1}{m_{\omega}^2 - Q_1^2 - i m_{\omega}\Gamma_{\omega}}
 \nonumber \\
 && {\kern-20pt}\left[ \ \theta\left((2.2 \ {\rm GeV})^2 - Q^2\right)
   + \theta\left(Q^2-(2.2 \ {\rm GeV})^2 \right)
 \left(\frac{\left(2.2 \ {\rm GeV}\right)^2}{Q^2}\right)^2 \right]
\end{eqnarray}

with 
\begin{eqnarray}
 m_{ \rho\prime} = 1.7 \ {\rm GeV} \ \  
 \Gamma_{\rho\prime} = 0.26 \ {\rm GeV} \ \ \sigma = -0.1 \nonumber \\
 m_{\omega} = 0.782 \ {\rm GeV} \ \ \Gamma_{\omega}= 0.0085 \ {\rm GeV} \ .
\end{eqnarray}       

 An additional suppression compared to \cite{Fink}  above
 \(Q^2 = \left(2.2 \ {\rm GeV}\right)^2\) was introduced.
 This modification prevents the unphysical growth of the cross section
 for very large \(Q^2\), which originates from the momentum dependent
 couplings in \Eq{om1}-\Eq{om2}.

\end{appendix}
\def\NP{{\sl Nuc. Phys.}} 
\def\PL{{\sl Phys. Lett.}} 
\def\PR{{\sl Phys. Rev.}} 
\def\PRL{{\sl Phys. Rev. Lett.}} 

\end{document}